\documentclass[useAMS,usenatbib]{mn2e}

\usepackage{graphicx}
\usepackage{xcolor}
\usepackage{amssymb, amsmath}
\usepackage{hyperref}
\hypersetup{colorlinks=true,citecolor=blue}
\usepackage[utf8]{inputenc}
\usepackage{amsmath, amssymb, bm}
\usepackage{multirow}

\renewcommand{\vec}[1]{\bmath{#1}}
\newcommand{\be}{\begin{equation}}
\newcommand{\ee}{\end{equation}}
\newcommand{\ba}{\begin{eqnarray}}
\newcommand{\ea}{\end{eqnarray}}

\newcommand{\nv}{\hat{\bf n}}
\newcommand{\kpar}{k_\parallel}

\title[BAO from 21cm intensity mapping with the SKA]
{Baryonic acoustic oscillations from 21cm intensity mapping: the Square Kilometre Array case}
\author[F. Villaescusa-Navarro, D. Alonso \& M. Viel]
{Francisco Villaescusa-Navarro$^{1,2}$\thanks{e-mail: villaescusa@oats.inaf.it}, 
David Alonso$^{3}$\thanks{e-mail: david.alonso@physics.ox.ac.uk}, 
Matteo Viel$^{1,2}$
 \\~\\
\footnotesize
\footnotesize 
$^1$ INAF, Osservatorio Astronomico di Trieste, via Tiepolo 11, I-34131 Trieste, Italy\\ 
$^2$ INFN -- National Institute for Nuclear Physics, Via Valerio 2, I-34127 Trieste, Italy\\ 
$^3$ University of Oxford, Denys Wilkinson Building, Keble Road, Oxford, OX1 3RH, UK\\
}

\begin{document}
\maketitle 

\begin{abstract}
  We quantitatively investigate the possibility of detecting baryonic acoustic oscillations (BAO) using
  single-dish 21cm intensity mapping observations in the post-reionization era. We show that the
  telescope beam smears out the isotropic BAO signature and, in the case of the Square Kilometer
  Array (SKA) instrument, makes it undetectable at redshifts $z\gtrsim1$. We however demonstrate that
  the BAO peak can still be detected in the radial 21cm power spectrum and describe a method to make
  this type of measurements. By means of numerical simulations, containing the 21cm cosmological
  signal as well as the most relevant Galactic and extra-Galactic foregrounds and basic instrumental
  effect, we quantify the precision with which the radial BAO scale can be measured in the 21cm
  power spectrum.
  
  We systematically investigate the signal-to-noise and the precision of the recovered BAO signal as
  a function of cosmic variance, instrumental noise, angular resolution and foreground contamination.
  We find that the expected noise levels of SKA would degrade the final BAO errors by $\sim5\%$ with
  respect to the cosmic-variance limited case at low redshifts, but that the effect grows up to
  $\sim65\%$ at $z\sim2-3$. Furthermore, we find that the radial BAO signature is robust against
  foreground systematics, and that the main effect is an increase of $\sim20\%$ in the final
  uncertainty on the standard ruler caused by the contribution of foreground residuals as well
  as the reduction in sky area needed to avoid high-foreground regions. We also find that
  it should be possible to detect the radial BAO signature with high significance in the full  redshift range.
  
  We conclude that a 21cm experiment carried out by the SKA should be able to make direct measurements
  of the expansion rate $H(z)$ with measure the expansion with competitive per-cent level precision on
  redshifts $z\lesssim2.5$.  
\end{abstract} 
 

\section{Introduction}
\label{sec:introduction}

The spatial distribution of matter in the Universe is sensitive to the value of the cosmological
parameters. Constraints on those can thus be placed by comparing the statistical properties of
the density field against predictions from theoretical models. Unfortunately, the true matter
density is not directly observable, and therefore one must resort to using proxies of it,
such as the number density of galaxies or line emission intensity of cosmic neutral hydrogen (HI). 

A promising and new way of tracing the large-scale structure of the Universe is to carry out
low angular resolution radio observations to detect the 21cm radiation from cosmic neutral
hydrogen in the post-reionization epoch. The idea is not to detect individual galaxies through
their 21cm emission, but rather to measure the combined flux in wide patches of the sky
containing many galaxies. This technique is called intensity mapping
\citep{Bharadwaj_2001A, Bharadwaj_2001B, Battye:2004re,McQuinn_2006, Chang_2008,
Loeb_Wyithe_2008, Villaescusa-Navarro_2014a, Bull_2015}. 
Under the assumption that the measured 21cm flux traces the perturbations in the matter
density on large-scales, we can use the clustering properties of the cosmic HI, as observed
from 21cm intensity mapping surveys, to put constraints on the value of the cosmological
parameters \citep{Bull_2015, Villaescusa-Navarro_2015a, Carucci_2015}. 

Baryonic acoustic oscillations (BAO), originated in the early Universe by the competition
between the gravitational interaction and the radiation pressure of photons tightly coupled
to baryons, leave an imprint in the late-time matter density in the form of a statistically
preferred separation between density peaks of $r_s\sim110~h^{-1}{\rm Mpc}$, corresponding
to the size of the sound horizon at the time of the baryon-photon decoupling. This translates
into a distinct peak in the matter/galaxy two-point correlation function, or as a set of
wiggles in the matter/galaxy power spectrum on scales $k\sim[0.05-0.3]~h{\rm Mpc}^{-1}$ with
frequency $r_s$. The BAO signature thus constitutes a cosmological standard ruler, whose
size depends on well understood physics of the early Universe. By measuring them in the
temperature anisotropies of the cosmic microwave background (CMB) and in the clustering
pattern of matter tracers, it is possible to measure the value of the Hubble rate and the
angular diameter distance as a function of redshift.

The main advantage of the BAO signal resides in its robustness against systematic effects:
it is difficult for non-cosmological effects to mimic or shift the position of
the BAO feature in the correlation function or power spectrum. Furthermore, given the large-scale
nature of the BAO signal, the effects induced by the non-linear gravitational evolution are
well captured by perturbation theory \cite[e.g.][]{Crocce_2008, Padmanabhan_2009,
Baldauf_2015,peloso}. 

The BAO scale has been measured in the 2pt/3pt statistics of galaxy surveys
\cite[see e.g.][]{Cole_2005, Eisenstein_2005, Anderson_2014, Gil-Marin_2015,Beutler_2016,
Alam_2016,Slepian_2016}, in the Ly$\alpha$-forest \citep{Delubac_2015}, in the distribution of
galaxy clusters \citep{alfonso} and in the spatial distribution of voids \citep{Kitaura_2016}. 
Upcoming and future radio experiments such as the Canadian Hydrogen Intensity Mapping
Experiment (CHIME)\footnote{\url{http://chime.phas.ubc.ca/}}, the Ooty Radio Telescope
(ORT)\footnote{\url{http://rac.ncra.tifr.res.in/}}, BINGO \cite{2012arXiv1209.1041B} and the
Square Kilometre Array (SKA)\footnote{\url{https://www.skatelescope.org/}} will survey large
areas of the sky using the intensity mapping technique in the post-reionization era. In this paper we
investigate the prospects of detecting the BAO from 21cm intensity mapping observations,
focusing on the SKA1-MID instrument. 

An ideal intensity mapping experiment would cover the largest possible field of view with
as large angular resolution as possible. Since the angular scales probed by a radio
interferometer are $\lambda/b_{\rm max}\lesssim\theta\lesssim\lambda/b_{\rm min}$, where $b_{\rm max/min}$
are the largest/smallest separation between two antenna elements, these two requirements
can only be simultaneously met by building large interferometric arrays of tightly packed
receivers. An alternative strategy would be to cover the desired sky footprint with
single-dish observations, in which case the angular resolution has a lower bound
$\theta\gtrsim\lambda/D_{\rm dish}$ determined by the dish diameter
\cite[see][for a detailed discussion]{2015ApJ...803...21B}. In this paper we will focus on
the latter case, the likely strategy of choice for the SKA1-MID instrument, described in
\cite{braun15}.

We demonstrate that the poor angular resolution inherent to single-dish 21cm observations
smears out the BAO peak in the isotropic correlation function or power spectrum, and that
in this case cosmological constraints would be driven by the overall shape of the 21cm
power spectrum, which is more sensitive to systematic effects. We will however show that
the BAO wiggles can be detected in the radial 21cm power spectrum, and thus can be
used to make a direct measurement of the expansion rate $H(z)$. In our analysis we will focus
on the impact of instrumental effects, such as the system noise, and the presence of
Galactic and extra-Galactic foregrounds on our the results. 

This paper is organized as follows. In Section \ref{sec:BAO} we study the impact of the
instrumental beam on the detectability of the isotropic BAO peak in single-dish experiments.
In Section \ref{sec:methods} we describe the simulations and analysis methods used in this
work. The results obtained from this analysis and their interpretation are presented in
Section \ref{sec:results}, where we systematically investigate the impact of each
complication (system noise and foregrounds) on the final uncertainties. Finally
we discuss main conclusions of this paper in Section \ref{sec:conclusions}.

\section{BAO and beam size}
\label{sec:BAO}

In this section we investigate the impact of the radio-telescope beam, when observations are
carried out using the single-dish technique, on the shape and position of the 21cm intensity
mapping BAO. We first analyze the possibility of detecting the isotropic BAO and then
we discuss the prospect of using the radial power spectrum to measure the BAO signal.

\subsection{Isotropic BAO}
\label{subsec:isotropic_BAO}

At linear order the 21cm power spectrum in real-space can be expressed as
\be
P_{\rm 21cm}(k,z)=b_{\rm 21cm}^2(z)P_{\rm m}(k,z)
\ee
where $b_{\rm 21cm}(z)=\bar{T}_b(z)b_{\rm HI}(z)$ is the bias of the 21cm signal and
$P_{\rm m}(k)$ is the linear matter power spectrum. $b_{\rm HI}(z)$ and $\bar{T}_b(z)$
are the is the HI bias and mean brightness temperature at redshift $z$, the latter
given by
\be
\bar{T}_b(z)=190\frac{H_0(1+z)^2}{H(z)}\Omega_{\rm HI}(z)h~~{\rm mK}~.
\ee
The telescope beam, which for simplicity we will model as being Gaussian, induces a
smoothing in the transverse direction 
\be
\delta_{\rm 21cm,obs}(\vec{k}_\bot,k_\|,z)=e^{-k_\bot^2R^2/2}\delta_{\rm 21cm}(\vec{k}_\bot,k_\|,z)
\ee
with $\delta_{\rm 21cm,obs}$, $\delta_{\rm 21cm}$ being the observed and cosmological 21cm
modes. The beam angular resolution $\theta_{\rm FWHM}$ is related to the transverse smoothing
scale as $R=r(z)\,\theta_{\rm FWHM}/(2\sqrt{2\ln2})$, where $r(z)$ is the
comoving angular diameter distance to redshift $z$. Note that given the frequency dependence
of the angular beam size, $\theta_{\rm FWHM}$ is implicitly also a function of $z$.

The observed 21cm power spectrum in real-space can thus be written as
\be
P_{\rm 21cm,obs}(k,\mu,z)=e^{-k^2R^2(1-\mu^2)}P_{\rm 21cm}(k,z)
\ee
and the monopole can be obtained by averaging over all modes sharing the value of
$k=\sqrt{k_\bot^2+k_\|^2}$:
\begin{eqnarray}
P_{\rm 21cm,obs}(k,z)&=&P_{\rm 21cm}(k,z)\frac{1}{2}\int_{-1}^1  e^{-k^2R^2(1-\mu^2)}d\mu\\
&=&\frac{{\mathcal{D}}{(kR)}}{kR}P_{\rm 21cm}(k,z)
\end{eqnarray}
where ${\mathcal{D}}(x)$ is the Dawson function. 
In configuration space the observed 21cm correlation function is given by
\be
\xi_{\rm 21cm,obs}(r,z)=\frac{b_{\rm 21cm}^2}{2\pi^2}\int_0^\infty k^2
P_{\rm m}(k,z)\frac{\sin(kr)}{kr}\frac{\mathcal{D}(kR)}{kR}dk
\ee
Note that so far we have neglected the effect of
redshift-space distortions. This is irrelevant for the main conclusions of this section, however 
their effect was fully taken into account in the simulations and theoretical models used in
Sections \ref{sec:methods} and \ref{sec:results}.

\begin{figure*}
  \centering
  \includegraphics[width=0.99\textwidth]{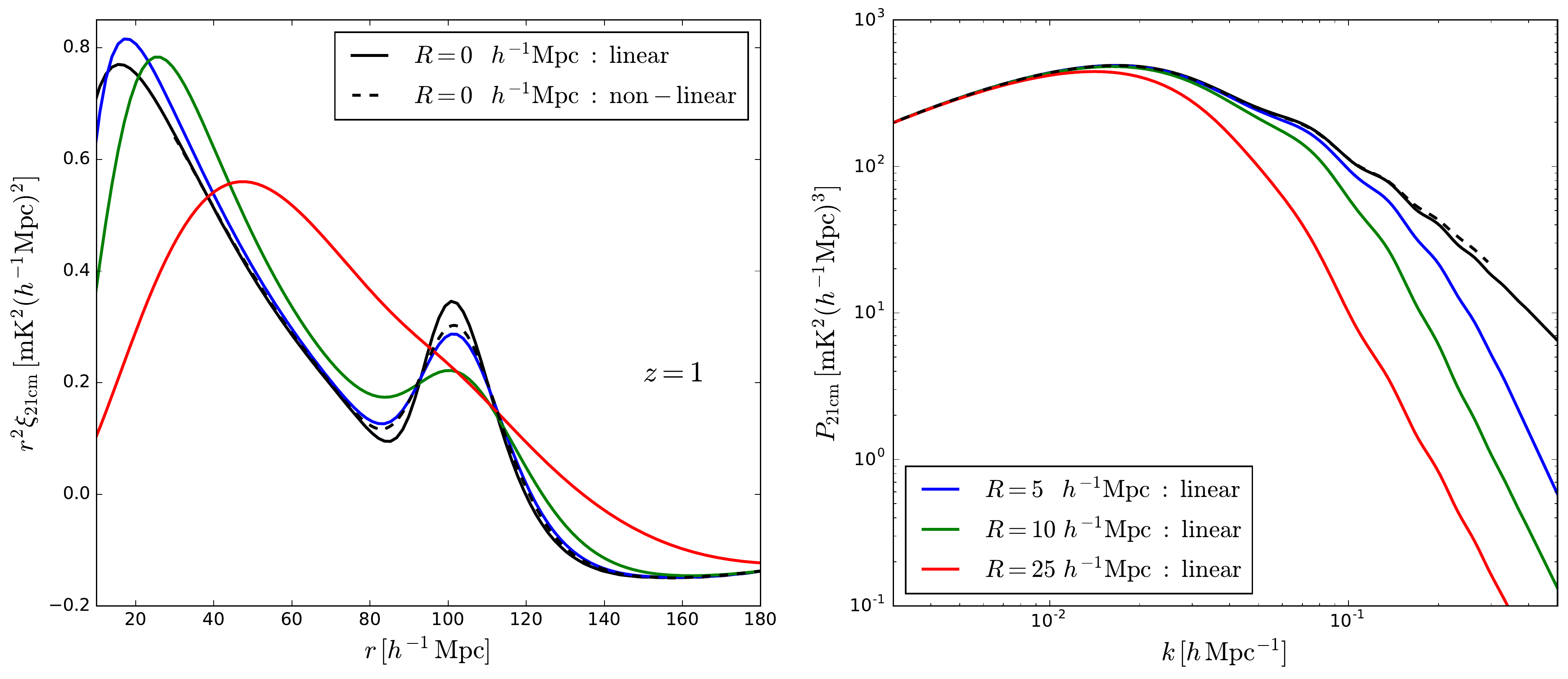}
  \caption{Impact of the telescope beam size on the isotropic BAO shape in configuration (left)
           and Fourier space (right) at $z=1$. The solid black line shows the results for
           infinite angular resolution, while the blue, green and red lines represent the
           observed correlation functions and power spectra for comoving smoothing scales
           $R=5$, 10 and 25 $h^{-1}$Mpc, respectively. The dashed black line displays the
           effect of non-linearities, computed using {\tt RegPT} at 2-loops for ininite
           angular resolution.}
  \label{fig:beam_BAO}
\end{figure*}

The left panel of Figure \ref{fig:beam_BAO} shows the observed 21cm correlation function at
$z=1$ for the fiducial cosmological model considered in this work (see Section
\ref{subsec:sims}) for different values of the angular smoothing scale $R$. Following our
reference HI model (described in Appendix \ref{sec:appendix_a}), we take $b_{\rm 21cm}=0.231$
mK at $z=1$. As can be seen in the figure,
in the idealized case of radio-telescopes having infinite resolution, the BAO peak can easily
be detected from single-dish IM observations. On the other hand, as the angular resolution of
the telescope decreases (either by going to higher redshift or by decreasing the antenna
diameter), the isotropic BAO peak is smeared out by the telescope beam. For angular smoothing
scales larger than $\sim 20~h^{-1}$Mpc the BAO peak is simply not visible. The Figure also
shows, with a dashed black line, the effect of non-linearities on the BAO peak at $z=1$ when
the 21cm maps have infinite angular resolution. The effect of non-linearities on the matter
power spectrum in real-space was computed using {\tt RegPT} at 2-loops\footnote{We have used
the {\tt RegPT} public code \url{http://www2.yukawa.kyoto-u.ac.jp/\~~atsushi.taruya/regpt\_code.html}.}.
As can be seen, even a relatively small angular smoothing like 5 $h^{-1}$Mpc has a larger impact
on the BAO signature than effects induced by non-linear gravitational evolution. The right panel
of Figure \ref{fig:beam_BAO} shows the analogous results in Fourier space.

It is useful to quantify the single-dish angular resolution that SKA1-MID will achieve as a
function of redshift. MID will consist of an array of 15 m antennae, corresponding to angular
resolutions of $\theta_{\rm FWHM}=\lambda/D\simeq0.8(1+z)$ deg. The corresponding comoving
smoothing scale will thus be given by $R=\{11.7, 27.2, 63.8, 104.1\}~h^{-1}$Mpc at redshifts
$z=\{0.5, 1.0, 2.0, 3.0 \}$, respectively. Thus, for redshifts $z\gtrsim1$ the poor angular
resolution of SKA1-MID will prevent the detection the isotropic BAO feature, and cosmological
constraints will be driven by the broadband shape of the 21cm power spectrum, which can be
significantly affected by systematic effects.

\subsection{Radial BAO}
\label{subsec:radial_BAO}

\begin{figure*}
  \centering
  \includegraphics[width=0.99\textwidth]{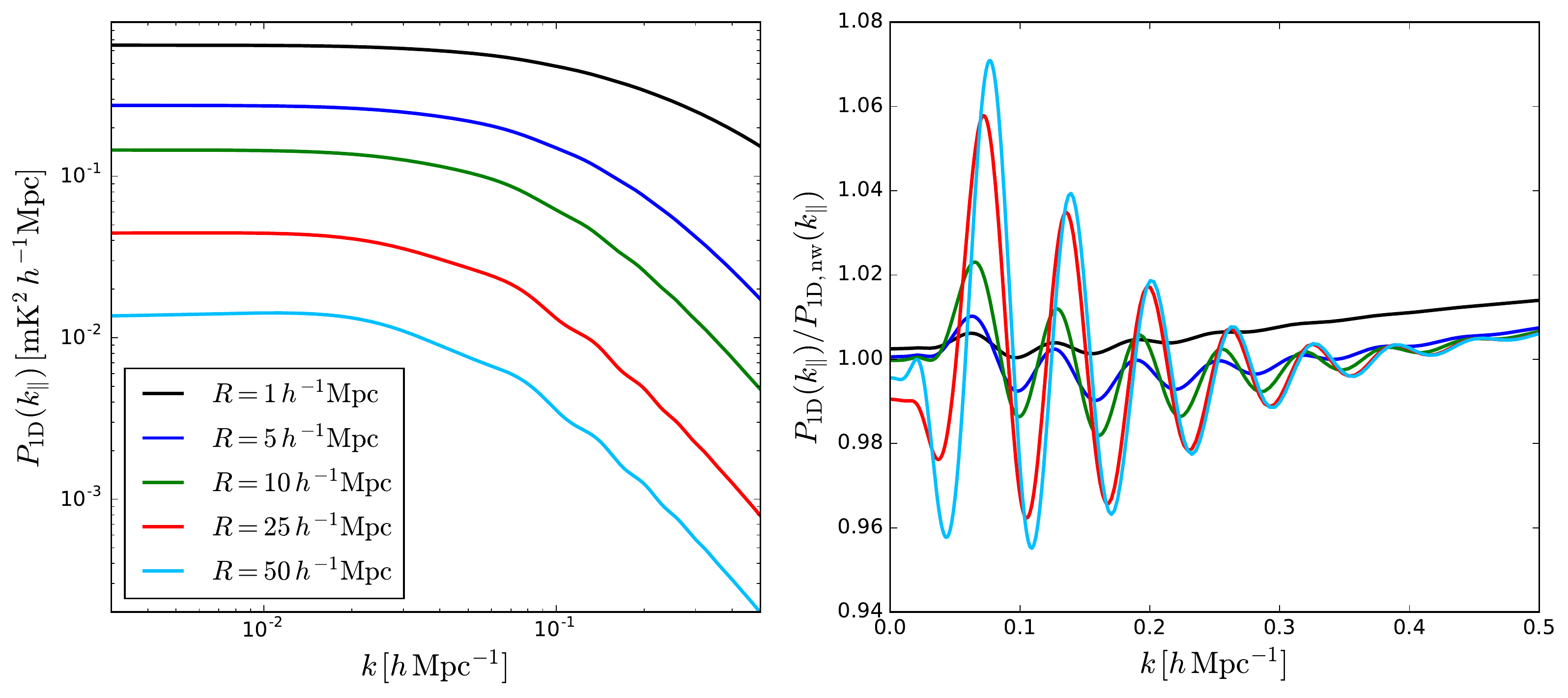}
  \caption{Impact of the telescope beam size on the BAO wiggles of the observed radial
           power spectrum. The left panel shows the radial power spectra at $z=1$ at
           linear order for different angular smoothing scales. In the right panel we
           show the results normalized by the radial 21cm power spectrum for a model
           with no BAO wiggles.}
  \label{fig:Pk_1D_beam}
\end{figure*}

In what follows we will define the 1D power spectrum $P_{\rm 1D}(k_\|)$ as
\begin{equation}
  \langle\delta_{\rm 1D}(k_\|,{\bf r}_\perp)\delta^*_{\rm 1D}(k_\|',{\bf r}_\perp)
  \rangle\equiv\delta^{\cal D}(k_\|-k_\|')\,P_{\rm 1D}(k_\|),
\end{equation}
where $\delta^{\cal D}$ is the Dirac $\delta$-function and $\delta_{\rm 1D}(k_\|,{\bf r}_\perp)$
is the one-dimensional Fourier transform along the radial direction of the overdensity field
for the transverse coordinate ${\bf r}_\perp$. It is straightforward to show that the relation
between the 1D and 3D power spectra is given by
\be
P_{\rm 1D}(k_\|,z)=\int \frac{d\vec{k}_\bot}{(2\pi)^2}P_{\rm 3D}(k_\|,\vec{k}_\bot,z)~.
\ee
In the limit of poor angular resolution (i.e. large $R$), it can be shown that 
\be
\lim\limits_{R \to \infty} \left(\frac{R^2}{\pi}e^{-k_\bot^2R^2}\right)=
\delta^{\cal D}(\vec{k_\bot})
\ee
and therefore we obtain
\be
\lim\limits_{R \to \infty} P_{\rm 21cm,obs,1D}(k_\|,z)=\frac{1}{4\pi R^2}P_{\rm 21cm}(k_\|,z)~.
\label{eq:1D_limit}
\ee
Thus, on the one hand the amplitude of the observed radial 21cm power spectrum scales inversely
proportional to the square of the angular smoothing scale. This is simply a consequence of the
supression of the contribution from perturbations on small transverse scales caused by the
large beam. On the other hand, for sufficiently large angular smoothing scales, the shape of the
observed 1D 21cm power spectrum will be the same as that of the 3D power spectrum in the absence
of instrumental beam\footnote{Note that in general the range of scales where this is a valid
approximation will depend on the smoothing scale. For $R=50~h^{-1}$Mpc both shapes match very
well up to $k\simeq0.01~h~{\rm Mpc}^{-1}$, but differ on larger scales.}. The consequence of
this is that the BAO wiggles can be easily identified in the observed radial 21cm power spectrum for
large angular smoothing scales.

These two effects can be seen in Fig. \ref{fig:Pk_1D_beam}, where we show, in the left panel,
the observed radial 21cm power spectrum for different angular smoothing scales. As expected, the
overall amplitude of the radial power spectrum decreases with increasing angular smoothing scale.
On the other hand, the BAO wiggles are more clearly visible when the angular smoothing scale is
larger. It is worth noting that the BAO wiggles are not visible in the case where the telescope
has a very large angular resolution. The reason for this is that the radial power spectrum at a
given wavenumber $k_\|$ receives contributions from all transverse scales larger than the
smoothing scale.
In the limit of zero angular smoothing, the large power on small scales significantly enhances
the amplitude the radial power spectrum, effectively decreasing the relative
amplitude of the BAO. The right panel of Figure \ref{fig:Pk_1D_beam} shows the ratio of the
1D power spectrum to the no-BAO power spectrum computed using the Eisenstein \& Hu fitting
formula \citep{EH_1998} for different angular smoothing scales. As can be seen, the BAO wiggles
are present in all cases, but they are more pronounced for large angular smoothing scales. We
however emphasize that this effect saturates once the limit of Eq. \ref{eq:1D_limit} is
achieved. By comparing the red and cyan lines
one can see that the relative amplitude of the BAO wiggles barely increases after doubling
the angular smoothing scale.

\section{Methods}
\label{sec:methods}

  In this section we describe the sky simulations, foreground-cleaning algorithm,
  radial power spectrum estimation method and the theoretical model we use to
  evaluate the detectability of the radial BAO scale.

\subsection{Simulations}
\label{subsec:sims}
\begin{figure*}
  \centering
  \includegraphics[width=0.49\textwidth]{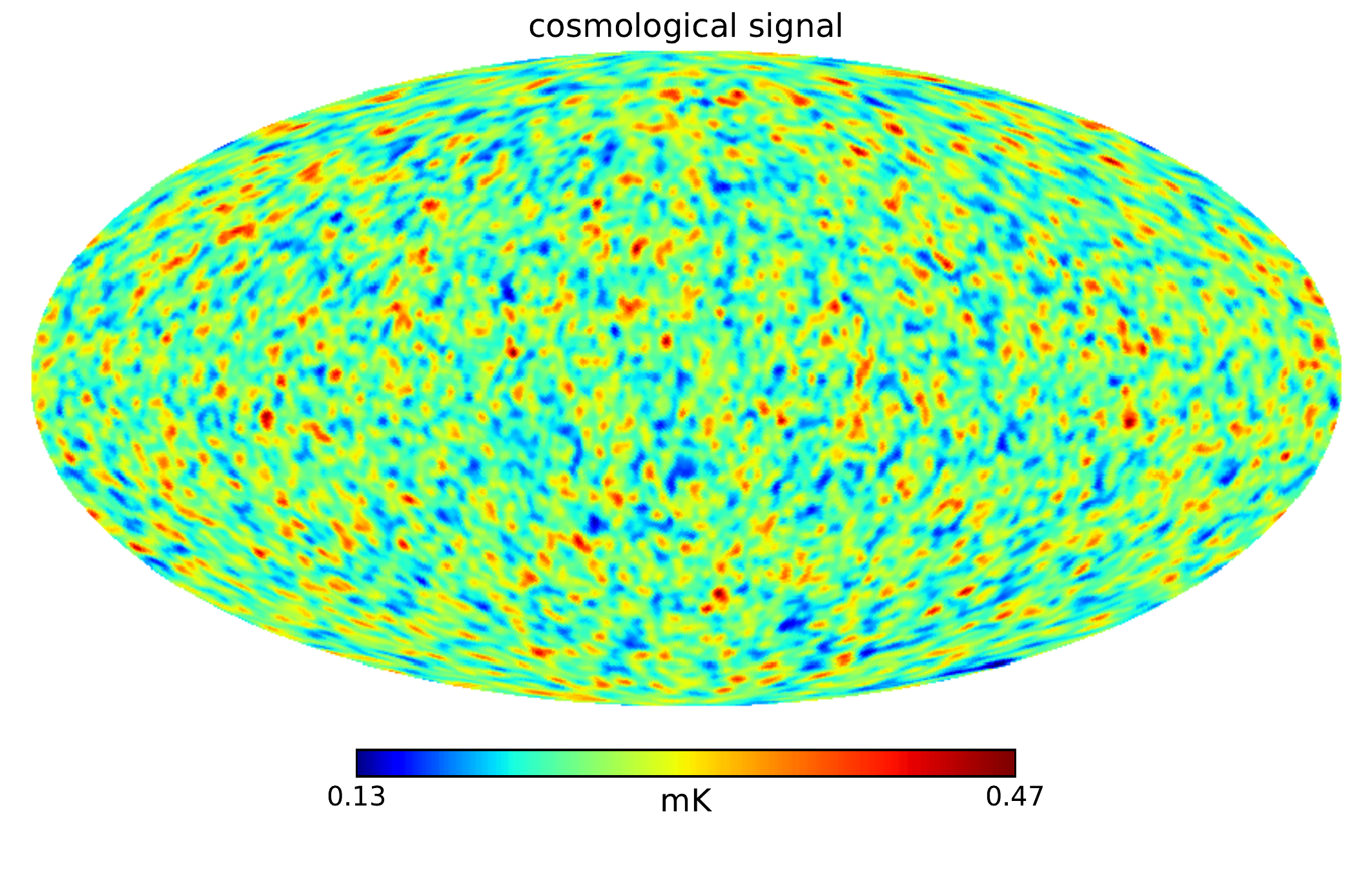}
  \includegraphics[width=0.49\textwidth]{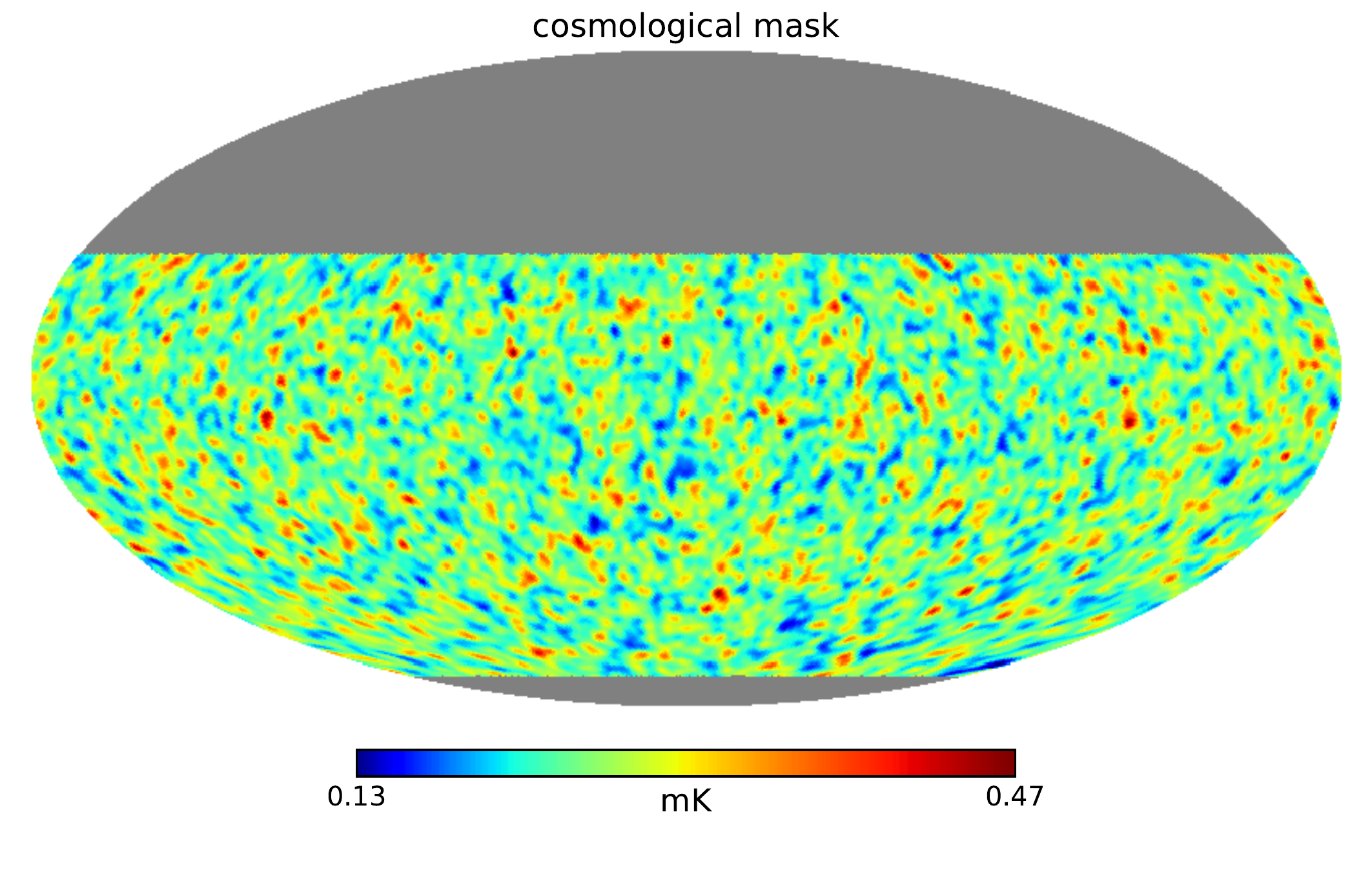}
  \includegraphics[width=0.49\textwidth]{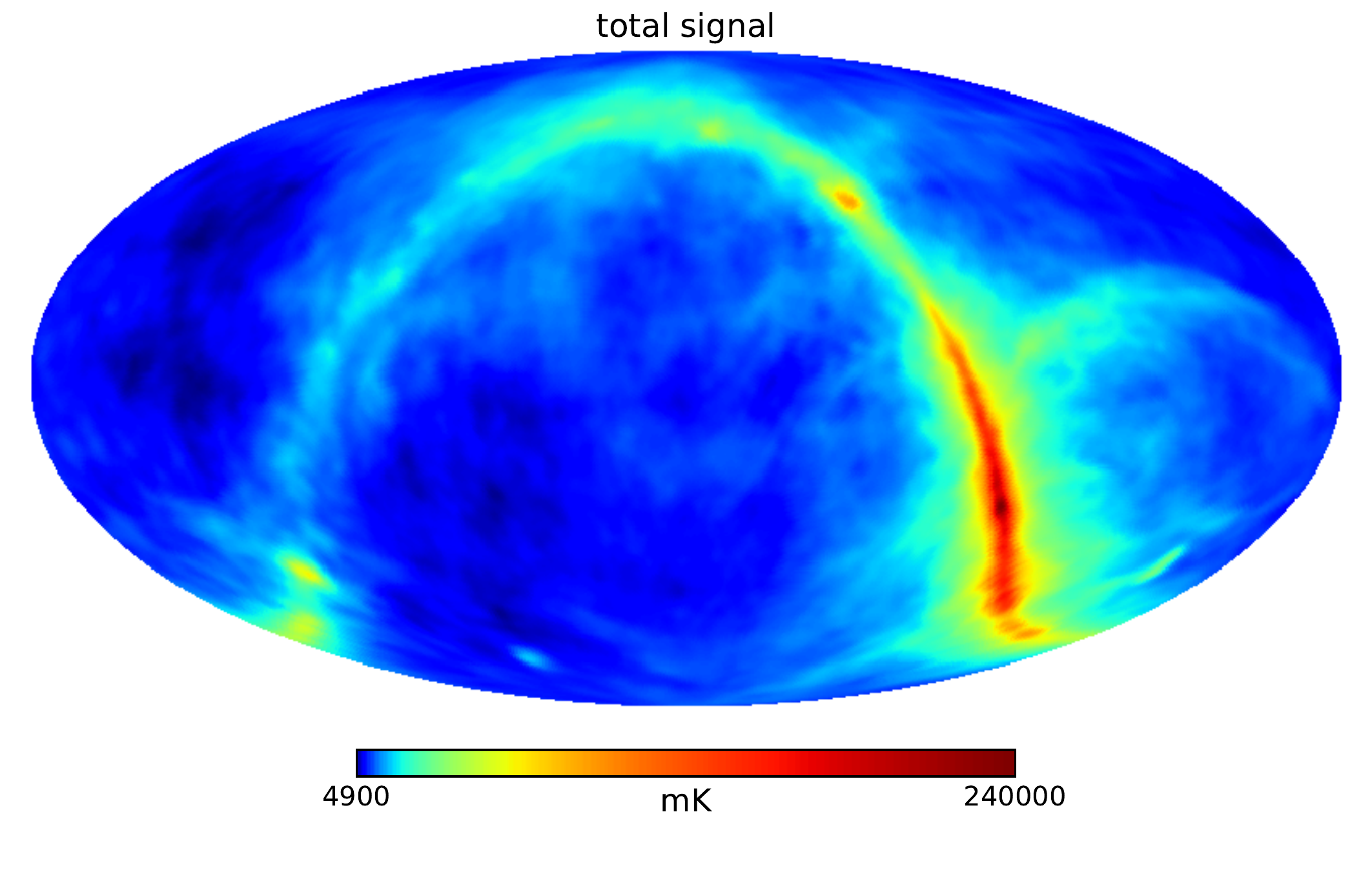}
  \includegraphics[width=0.49\textwidth]{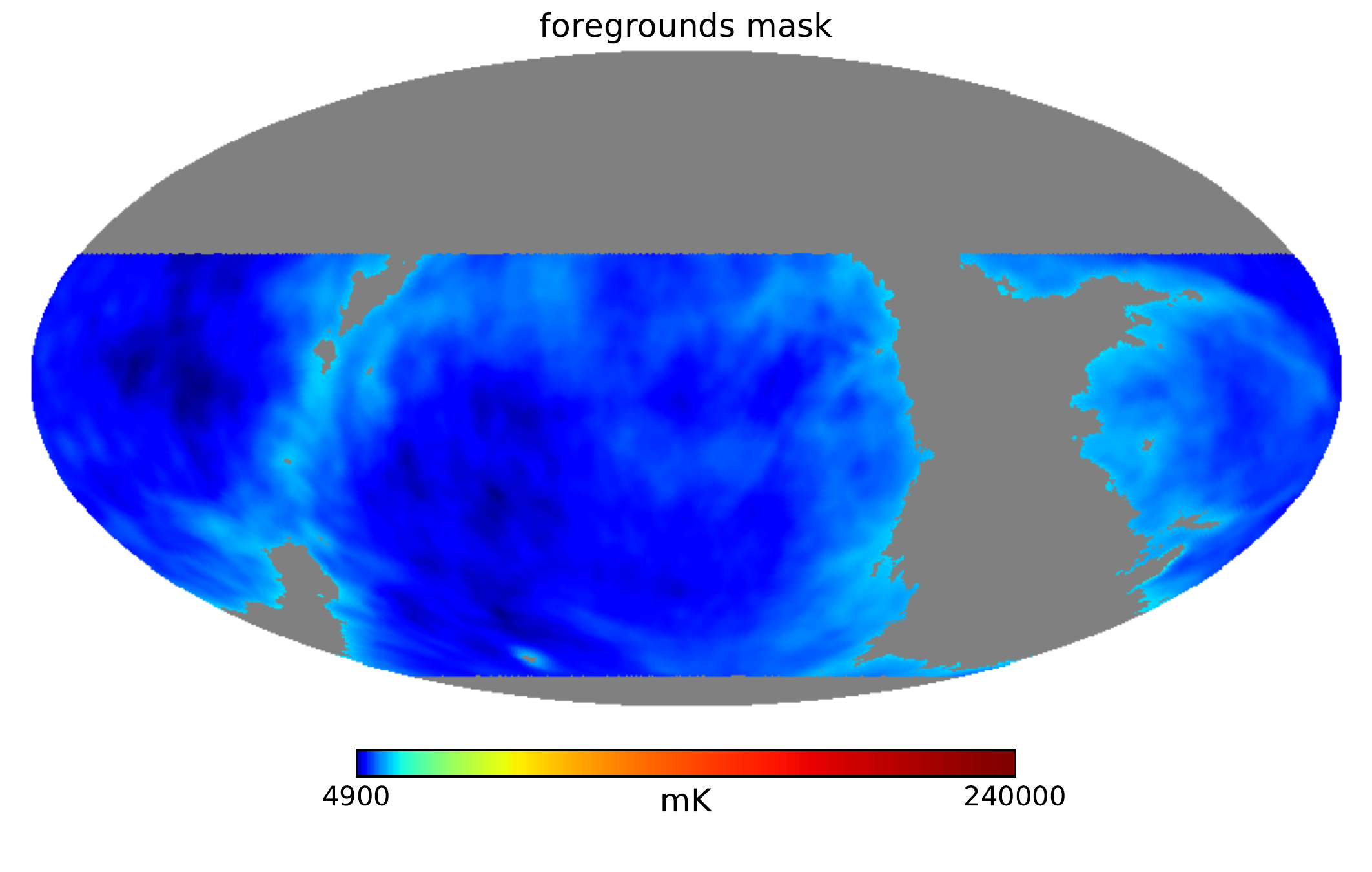}
  \caption{Maps at $\nu=541$ MHz ($z=1.62$) in equatorial coordinates containing the cosmological
           HI signal and {\bf noise only} (upper row) and including the contribution from Galactic and
           extra-Galactic foregrounds, (bottom row). The masks we use in our analysis
           are shown in the right column. The mask used for simulations without foregrounds
           is shown in the upper-right panel (the \textcolor{red}{\textit{cosmological mask}}) is defined by a simple
           cut in declination compatible with SKA the observing site.
           The bottom-right panel displays the mask we employ for maps containing foregrounds (the
           \textcolor{blue}{\textit{foregrounds mask}}): beyond the declination cuts, we remove regions of high
           foreground emission.}
  \label{fig:mask}
\end{figure*}
  We used the public code described in \cite{2014MNRAS.444.3183A} to generate a suite of
  100 simulations of the 21cm intensity mapping signal, as well as the main sources of
  foreground contamination. We describe here the most important components of these
  simulations, and refer the reader to \cite{2014MNRAS.444.3183A} for further details.
  \begin{itemize}
    \item {\bf Cosmological signal.} The code uses a simplified lognormal model to relate
          the cosmological HI intensity to the underlying dark matter density. The
          method involves generating a Gaussian realization of the linear density
          and velocity fields in a Cartesian grid at redshift $z=0$ assuming a given
          model for the matter power spectrum. Lightcone evolution is then
          implemented by placing the observer at the centre of the simulation box
          and assuming linear growth and a purely redshift-dependent clustering
          bias $b_{\rm HI}$. The density field is then subjected to a local
          lognormal transformation and put in redshift space using the Gaussian
          velocity field. Finally, the Cartesian HI overdensity field is interpolated
          onto a set of sky maps at different frequencies. These are defined using
          the HEALPix pixelization scheme \cite{2005ApJ...622..759G}, and the total HI
          temperature in each pixel is computed assuming a model for the background HI
          density $\Omega_{\rm HI}$.
          
          Our simulations were generated using a box of $8850\,h^{-1}{\rm Gpc}$
          on a side with $3072^3$ grid cells, enough to produce temperature maps
          in the range of frequencies $\nu=[350,1050]\,{\rm MHz}$. The sky maps were 
          generated using a HEALPix resolution parameter $N_{\rm side}=256$,
          corresponding to a resolution of $\theta_{\rm pix}\sim14'$, significantly
          better than the angular resolution achievable with a single-dish intensity
          mapping experiment carried out with SKA1-MID. The density and velocity
          fields were generated for a matter power spectrum corresponding to the
          best-fit flat $\Lambda$CDM cosmological parameters found by
          \cite{2014A&A...571A..16P}:
          $(\Omega_M,\Omega_b,h,n_s,\sigma_8)=(0.315,0.049,0.67,0.96,0.83)$. The
          HI temperature anisotropies were generated assuming the models described
          in Appendix \ref{sec:appendix_a}:
          \begin{equation}
            \Omega_{\rm HI}(z)=\Omega_{\rm HI,0}\,(1+z)^\alpha,\hspace{12pt}
            b_{\rm HI}(z)=b_{\rm HI,0}+b'\,(1+z)^\beta,
          \end{equation}
          where $\Omega_{\rm HI,0}=4\times10^{-4},$ $\alpha=0.6,$ $b_{\rm HI,0}=0.904$,
          $b'=0.135$ and $\beta=1.70$.
          
    \item {\bf Foregrounds.} The foreground simulations are based on the models of
          \cite{2005ApJ...625..575S}. The angular fluctuations for each foreground
          component as a function of frequency are simulated as Gaussian random 
          realizations of an angular-frequency power spectrum parametrised as:
          \begin{equation}
            C_\ell(\nu_1,\nu_2)=A\left(\frac{\ell_{\rm ref}}{\ell}\right)^\beta
            \left(\frac{\nu_{\rm ref}^2}{\nu_1\nu_2}\right)^\alpha
            \exp\left[-\frac{\log^2(\nu_1/\nu_2)}{2\xi^2}\right],
          \end{equation}
          where $\xi$ corresponds to the correlation length in frequency space. Thus,
          in the limit $\xi\rightarrow0$ the foregrounds are perfectly correlated
          ($C_\ell(\nu_1,\nu_2)/\sqrt{C_\ell(\nu_1,\nu_1)C_\ell(\nu_2,\nu_2)}=1$),
          which corresponds to the simplest case in terms of foreground removal.

          We simulate four foreground components: Galactic synchrotron, extragalactic
          point sources, Galactic and extra-Galactic free-free emission. The values of
          the parameters $A,\,\beta,\,\alpha$ and $\xi$ are given in Table 1 of
          \cite{2014MNRAS.444.3183A}. In the case of Galactic synchrotron we also simulate
          the large-scale Galactic emission by extrapolating the Haslam map
          \cite{1982A&AS...47....1H} at 408 MHz to other frequencies. Fluctuations on
          scales smaller than those probed by the  Haslam map ($\sim1^\circ$), as well
          as frequency decorrelation, are added using the model above.

    \item {\bf Instrument.} As our baseline intensity mapping experiment we use the
          first phase of the SKA1-MID array, consisting of $N_{\rm dish}\simeq200$,
          $D_{\rm dish}=15\,{\rm m}$ dishes with an instrument temperature
          $T_{\rm inst}=25\,{\rm K}$. The instrumental noise was simulated as white, Gaussian
          noise, with a variance per steradian given by
          \begin{equation}
            \sigma_N^2(\nu)=
            T_{\rm sys}^2(\nu)\frac{4\pi\,f_{\rm sky}}{N_{\rm dish}\,t_{\rm tot}\,\Delta\nu},
          \end{equation}
          where the system temperature is $T_{\rm sys}=T_{\rm inst}+(60\,{\rm K})
          (\nu/300 {\rm MHz})^{-2.5}$, and we assumed a total observation time of
          $t_{\rm tot}=10000\,{\rm h}$.
          
          Finally, the instrumental beam was simulated as being Gaussian, with a width
          $\theta_{\rm FWHM}=\lambda/D_{\rm dish}$. This assumes single-dish observations,
          which corresponds to the most optimal use of SKA1 as an intensity mapping
          experiment for cosmological purposes \cite{2015ApJ...803...21B,2015aska.confE..19S}.          
    \item {\bf Mask.} SKA1-MID will be physically located in South Africa. Therefore, it can
          not make full-sky observations. We defined the expected field of view for SKA assuming
          the maximum observable area, corresponding to a range in declination
          ${\rm dec}\in(-75^\circ,28^\circ)$. In what follows we will label the mask corresponding to
          this field of view the {\sl cosmological mask}, and we will use it to study simulations
          in the absence of foregrounds. Besides this cut in declination we also defined a Galactic
          mask by removing all pixels with synchrotron emission above $40\,{\rm K}$ at
          $408\,{\rm MHz}$. This removes approximately $20\%$ of the observable sky, and was found
          to be an optimal compromise between sky coverage and foreground residuals by
          \citet{2015MNRAS.447..400A}. We will label the mask resulting from the combination of the
          declination and synchrotron cuts the {\sl foregrounds mask}, and we will employ it when
          studying simulations with foregrounds. The sky fractions of the cosmological and foregrounds
          mask are $f_{\rm sky}=0.72$ and $0.58$ respectively.          
  \end{itemize}
  
  For each simulation we generated 691 sky maps, each covering a frequency band corresponding
  to a constant comoving radial separation $\Delta\chi\equiv\Delta\nu(1+z)^2/H(z)=
  5\,h^{-1}{\rm Mpc}$. We produced 3 different types of maps: 1) maps containing only the cosmological
  signal, 2) maps containing the cosmological signal and instrument noise, 3) maps containing the cosmological
  signal, system noise and Galactic and extra-Galactic foregrounds. In all maps we take into account
  the instrument beam as described above. Figure \ref{fig:mask} illustrates the different components included
  in the simulations, as well as the sky masks used in the analysis.

\subsection{Foreground removal}
\label{subsec:fg}
  In simulations containing foregrounds, we applied a blind foreground cleaning
  algorithm to every simulation. For this we used the public code described
  in \cite{2015MNRAS.447..400A}, in particular applying a principal component analysis method
  (PCA). This algorithm follows three steps:
  \begin{enumerate}
    \item We estimate the frequency-frequency inverse-variance weighed covariance matrix by
          averaging over all available $N_{\rm pix}$ pixels:
          \begin{equation}
            C_{ij}=N_{\rm pix}^{-1}\sum_{n=1}^{N_{\rm pix}}
                   \frac{T(\nu_i,\nv_n)}{\sigma_i}
                   \frac{T(\nu_j,\nv_n)}{\sigma_j}.
          \end{equation}
          Here, $\sigma_i$ is the standard deviation of the frequency-decorrelated components
          in the $i_{\rm th}$ frequency channel, which ideally should receive contributions
          from both the instrumental noise and the intensity mapping signal.
    \item The covariance matrix is diagonalised:
          \begin{equation}
            \hat{\sf U}^T\hat{\sf C}\,\hat{\sf U}={\rm diag}(\lambda_1,...,\lambda_{691}),
          \end{equation}
          and the eigenvalues arranged in descending order ($\lambda_i>\lambda_{i+1}$).
    \item The first $N_{\rm fg}$ largest eigenvalues are then identified as encoding the
          main foreground contribution, and the foreground-clean maps are generated by
          subtracting all modes corresponding to the eigenvectors of these eigenvalues.
          The number of foreground modes to subtract, $N_{\rm fg}$, was determined as a
          compromise between foreground contamination and signal loss. This will be
          described in more detail in Section \ref{subsec:foregrounds}.
  \end{enumerate}

  \subsection{The 1D power spectrum}\label{subsec:pk}
  The method proposed in this work to measure the radial BAO scale is based on measuring
  the 1-dimensional radial 21cm power spectrum. Let us start by considering the HI temperature
  fluctuations Fourier-transformed along one particular line of sight $\nv$:
  \begin{equation}
    \Delta T(\kpar,{\bf r}_\perp)\equiv \int\frac{dr_\parallel}{\sqrt{2\pi}}
    \Delta T(r_\parallel,{\bf r}_\perp)\,e^{i\kpar r_\parallel},
  \end{equation}
   where we have made use of the flat-sky approximation, relating the observable quantities,
  frequency $\nu$ and angular coordinates $\nv$, with cartesian coordinates $r_\parallel$
  and ${\bf r}_\parallel$ through:
  \begin{equation}
    r_\parallel\equiv\chi(z),\hspace{6pt}{\bf r}_\perp\equiv\chi(z)\nv,
  \end{equation}
  where $\chi$ is the radial comoving distance to redshift $z\equiv\nu/\nu_{21}-1$.  
  It is then easy to prove that the two-point function of this observable is given by
  \begin{equation}
    \left\langle\Delta T(\kpar,{\bf r}_\perp)\Delta T^*(\kpar',{\bf r}_\perp')\right\rangle
    \equiv \delta(\kpar-\kpar')P_\parallel(\kpar,|{\bf r}_\perp-{\bf r}_\perp'|),
  \end{equation}
  where
  \begin{equation}
    P_\parallel(\kpar,\sigma)=\int_0^\infty\frac{dk_\perp\,k_\perp}{2\pi}J_0(k_\perp\sigma)
    W^2_p(k_\perp)P_T(\kpar,k_\perp).
  \end{equation}
  Here $J_0(x)$ is the 0-th order cylindrical Bessel function, $W_p$ is the pixel window
  function and $P_T(\kpar,k_\perp)$ is the three-dimensional power spectrum of the
  HI temperature fluctuations. Finally, we define the 1D radial power spectrum as the
  two-point function above for zero angular separation:
  \begin{equation}
    P_{\rm 1D}(\kpar)\equiv P_\parallel(\kpar,\sigma=0).
  \end{equation}
  
  Following this logic, we estimate the 1D power spectrum from the simulations through
  the following process:
  \begin{enumerate}
    \item We start by dividing the full frequency range of our simulations into a number
          of wide frequency bins. The width of these bins should be chosen such that
          the radial BAO scale can be sufficiently well sampled. We thus chose to use
          the 4 frequency bins described in Table \ref{tbl:survey}, corresponding to
          roughly equivalent comoving radial separations. Note that, given the
          frequency-dependent beam size, we reduced the pixel resolution of the intensity
          maps to $N_{\rm side}=64$ in the first 3 bins, and $N_{\rm side}=32$ in the
          lower-frequency bin.
    \item Within each bin, we compute the Fast Fourier Transform (FFT) of each pixel
          individually, and then compute the radial power spectrum for each pixel as
          the modulus of this FFT.
    \item Finally, we average this power spectrum over all pixels in the observed sky
          region.
  \end{enumerate}
  
  Note that this estimation of the power spectrum assumes that many effects remain
  constant inside the frequency bin, such as the growth of perturbations, the background
  HI temperature or the comoving scale corresponding to the instrumental beam. The
  smooth frequency/redshift dependence of these effects, however, should introduce
  broadband modifications in the estimated 1D power spectrum, which must be accounted
  for when measuring the BAO scale. We will study these effect further in Section
  \ref{subsec:fit}.

\subsection{Fitting}
\label{subsec:fit}

In order to derive constraints on the value of the cosmological and astrophysical
parameters we need a theoretical model to explain the observed or simulated data.
The theoretical template we use to model the shape and amplitude of the radial
21cm power spectrum is:
\begin{align}\nonumber
  P_{\rm model}(k_\|,z|\vec{\Theta})=&\left[P_{\rm lin,1D}(k_\|/\alpha,z)-
  P_{\rm nw,1D}(k_\|/\alpha,z)\right]e^{-k_\|^2\Sigma^2}\\\label{template}
                                     &+P_{\rm nw,1D}(k_\|,z)+A(k_\|)
\end{align}
with $\Theta$ a set of parameters to be defined below and where $P_{\rm lin,1D}(k)$ and
$P_{\rm nw,1D}(k)$ represent the 1D linear power spectra with and without BAO wiggles
respectively \citep{EH_1998}, computed from their 3D counterparts as
\be
P_{\rm i,1D}(k_\|)=\int \frac{d^2\vec{k}_\bot}{(2\pi)^2}P_{\rm i,3D}(k_\|,\vec{k}_\bot)~.
\ee
Here we shall model the 3D power spectrum as
\be
P_{\rm 21cm}(k_\|,\vec{k}_\bot)=b_{\rm fit}^2\left(1+\beta \mu^2\right)^2e^{-(k_\bot R)^2}P_{\rm m}(k),
\ee
which contains the effect of linear redshift-space distortions as in \citet{Kaiser_1987}, with $\mu\equiv k_\|/k$
and $\beta\equiv f(z)/b_{\rm HI}(z)$, where $f(z)$ is the linear growth rate. The exponential term models
the smoothing in the transverse direction induced by the beam of the instrument. Since we measure the 1D
power spectrum along maps at different frequencies, and therefore with different angular smoothing, we
introduce the parameter $R$ to model the \textit{effective} transverse smoothing scale in the power
spectrum measurement. $R$ represents thus a nuisance parameter whose value we marginalize over when
deriving constraints on the cosmological and astrophysical parameters.

The parameter $\Sigma$ in the equations above controls the damping and broadening of the BAO peak
induced by non-linear effects. Since our simulations are not able to fully capture the non-linear
gravitational effects, and since the amplitude of this parameter is expected to be small at high
redshifts, we fix $\Sigma=0$ when fitting the results of the simulations using the
theoretical template.
\begin{table}
\centering
\begin{tabular}{ccccc}
\hline \hline
$\nu$ $({\rm MHz})$ & $z$ & $\langle z\rangle$ & Vol. $(h^{-1}\,{\rm Gpc})^3$ & $N_{\rm side}$\\
\hline
$[812-1044]$ & $[0.36-0.75]$ & 0.6 & (\textcolor{gray}{22}, \textcolor{red}{16}, \textcolor{blue}{13}) & 64\\
\hline
$[627-812]$  & $[0.75-1.26]$ & 1.0 & (\textcolor{gray}{56}, \textcolor{red}{40}, \textcolor{blue}{32}) & 64\\
\hline
$[476-627]$  & $[1.26-1.98]$ & 1.6 & (\textcolor{gray}{107}, \textcolor{red}{77}, \textcolor{blue}{62}) & 64\\
\hline
$[350-476]$  & $[1.98-3.05]$ & 2.5 & (\textcolor{gray}{172}, \textcolor{red}{123}, \textcolor{blue}{99}) & 32\\
\hline \hline
\end{tabular}
\caption{Characteristics of the four redshift bins used in our analysis. The first and second columns
         show the frequency and redshift range, while the third column displays the mean redshift of the
         bin. The fourth column shows the comoving volume covered by each redshift bin, with the numbers
         displayed in \textcolor{gray}{gray}, \textcolor{red}{red} and \textcolor{blue}{blue} corresponding
         to simulations using \textcolor{gray}{no mask}, using the \textcolor{red}{cosmological mask} and
         using the \textcolor{blue}{foregrounds mask} respectively. The fifth column shows the HEALPix resolution
         parameter $N_{\rm side}$ us in the analysis of each bin. The radial widths and radial
         resolutions of all bins are constant and correspond to $0.9\,h^{-1}{\rm Gpc}$ and
         $5\,h^{-1}{\rm Mpc}$ respectively.} \vspace{-1.5em}
\label{tbl:survey}
\end{table}

The shape of the power spectrum can be modified by effects such as foreground-removal biases,
the frequency-dependent beam, evolution effects within the redshift bin, scale- and
redshift-dependent bias or non-linearities; we can therefore anticipate differences
between the simplified model described so far for the power spectrum and the actual
measurements. All of these effects, however, should only give a broadband contribution
to the observed power spectrum, and therefore we attempt to account for them using the
polynomial $A(k)=p_0k+p_1+p_2/k$, where the coefficients are treated as free parameters
in the fit and are marginalized out.

In summary, our model for the radial 21cm power spectrum contains 6 free parameters
\be
\vec{\Theta}=\{\alpha,b_{\rm fit},R,p_0,p_1,p_2\}~.
\ee
The cosmological information is encoded in the value of $\alpha$, which can be related
to the expansion rate as $H(z)=H_{\rm fid}(z)/\alpha$, where $H_{\rm fid}$ is the
expansion rate in the fiducial cosmology chosen to analyze the data. Thus, under the
assumption that the fiducial cosmology is close enough to the real one, the measurements
of $\alpha$ should be compatible with $1$. Note that we have included the overall
amplitude of the power spectrum, encoded in the effective bias $b_{\rm fit}$, as a
free parameter of the model. $R$ and the polynomial coefficients, $p_0, p_1, p_2$, are
nuisance parameters that model the instrument beam and the broad-band shape of the
power spectrum respectively.

Given measurements of the 1D power spectrum in a number of bins in the wavenumbers $k_\|$
at redshift $z$, that we refer to as the \textit{data} $\mathcal{D}$, we can place
constraints on the model parameters by exploring their posterior distribution, given
by the data likelihood via Bayes theorem. Assuming that the measurements of the
power spectrum are Gaussianly distributed, we can write
\be
-2\log\left[\mathcal{P}(\vec{\Theta}|\mathcal{D})\right]=
(\vec{P}-\vec{P}_{\rm model})^T\hat{\sf C}^{-1}(\vec{P}-\vec{P}_{\rm model}),
\ee
where $\hat{\sf C}$ is the covariance matrix and $\vec{P}$ and $\vec{P}_{\rm model}$ are the
measurements and model of the 1D power spectrum. We estimated the covariance matrix from
our 100 simulations and we found it to be effectively diagonal (see Appendix
\ref{sec:appendixB}), as expected on linear and mildly non-linear scales. Thus, in our
analysis we will assume a diagonal covariance matrix, which is a valid approximation
on the scales relevant for this paper.

The best fit, confidence levels and correlations between parameters were obtained by
exploring the parameter space with a Monte Carlo Markov Chain (MCMC) method using
the publicly available code {\tt emcee} \citep{emcee}. The quality of the fit was
quantified through the value of $\chi^2\equiv-2\log\mathcal{P}(\vec{\Theta}|\mathcal{D})$.

\section{Results}
\label{sec:results}

In this section we present the results of our analysis in terms of
constraints on the value of the BAO scaling parameter $\alpha$.
In order to isolate and understand the effects of different processes
affecting the measured signal, we carry out our analysis over three
different types of 21cm maps: 1) maps containing only the cosmological
signal, 2) maps with the cosmological signal plus instrument noise and
3) maps with the cosmological signal, noise from the instrument and
residual temperature fluctuations arising by the imperfect cleaning
of the foregrounds. Finally, we quantify the significance of the BAO
features on the data.

\subsection{Redshift binning and S/N ratio}
\label{subsec:binning}

As discussed above, each simulation consists of 691 maps equally spaced
in radial comoving distance from $z=0.35$ to $z=3.05$. As a compromise
between comoving volume, sampling rate of the BAO scale and the need
to capture the redshift evolution of the expansion rate, we split our
maps into the four redshift bins summarized in Table \ref{tbl:survey}.
We chose these redshifts bins to have equal radial comoving width,
rather than equal volume, so that the same number of radial modes would
be sampled in all of them.

\begin{figure}
  \centering
  \includegraphics[width=0.49\textwidth]{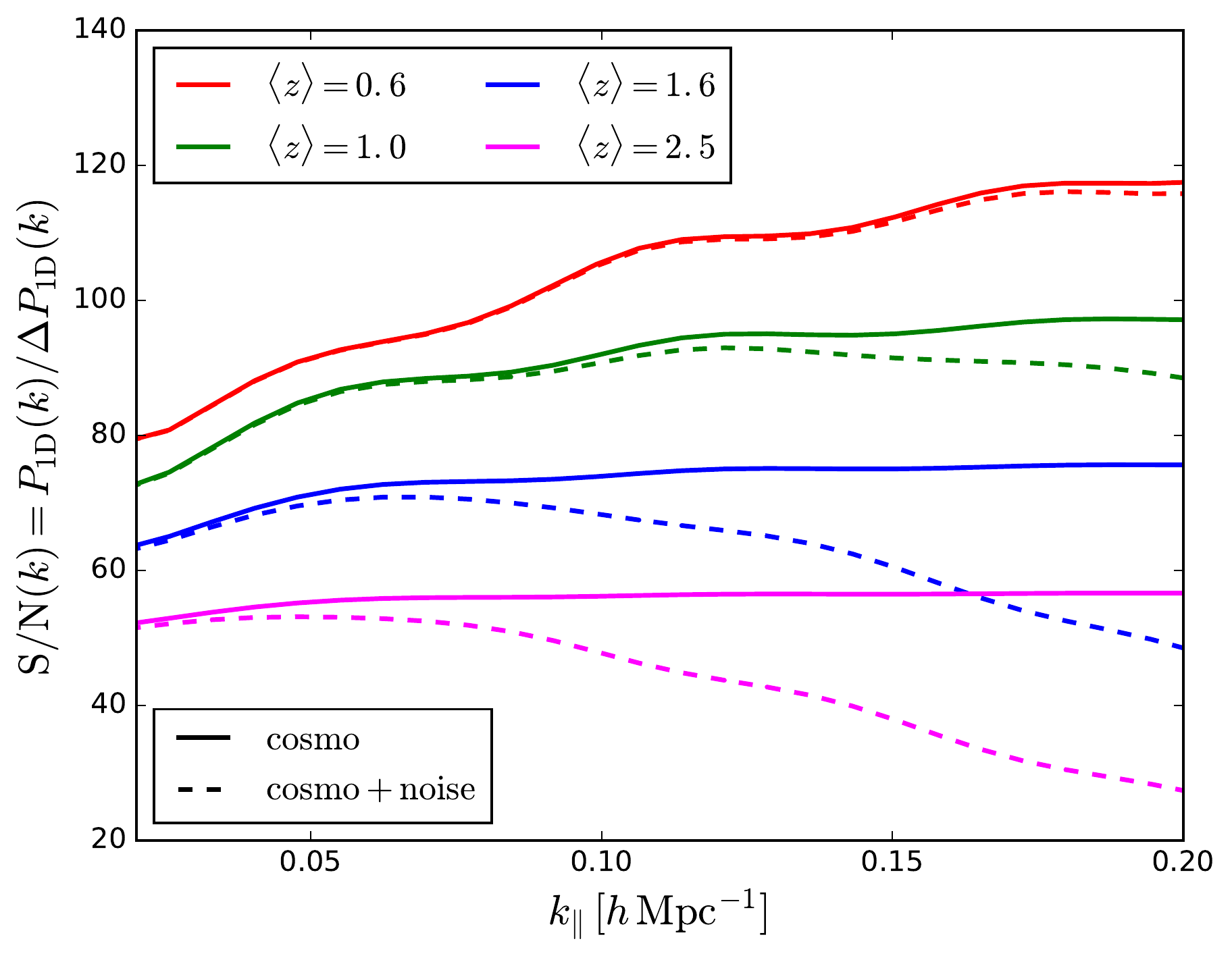}
  \caption{Signal-to-noise ($S/N$) ratio of the radial 21cm power spectrum in our fiducial redshift bins
           for maps containing only the cosmological signal (solid lines) and the cosmological signal
           plus system noise (dashed lines) using the cosmological mask. The $S/N$ ratio is defined as
           the ratio between the amplitude and the error of the radial power spectrum.}
  \label{fig:signal-to-noise}
\end{figure}

\begin{figure*}
  \centering
  \includegraphics[width=0.95\textwidth]{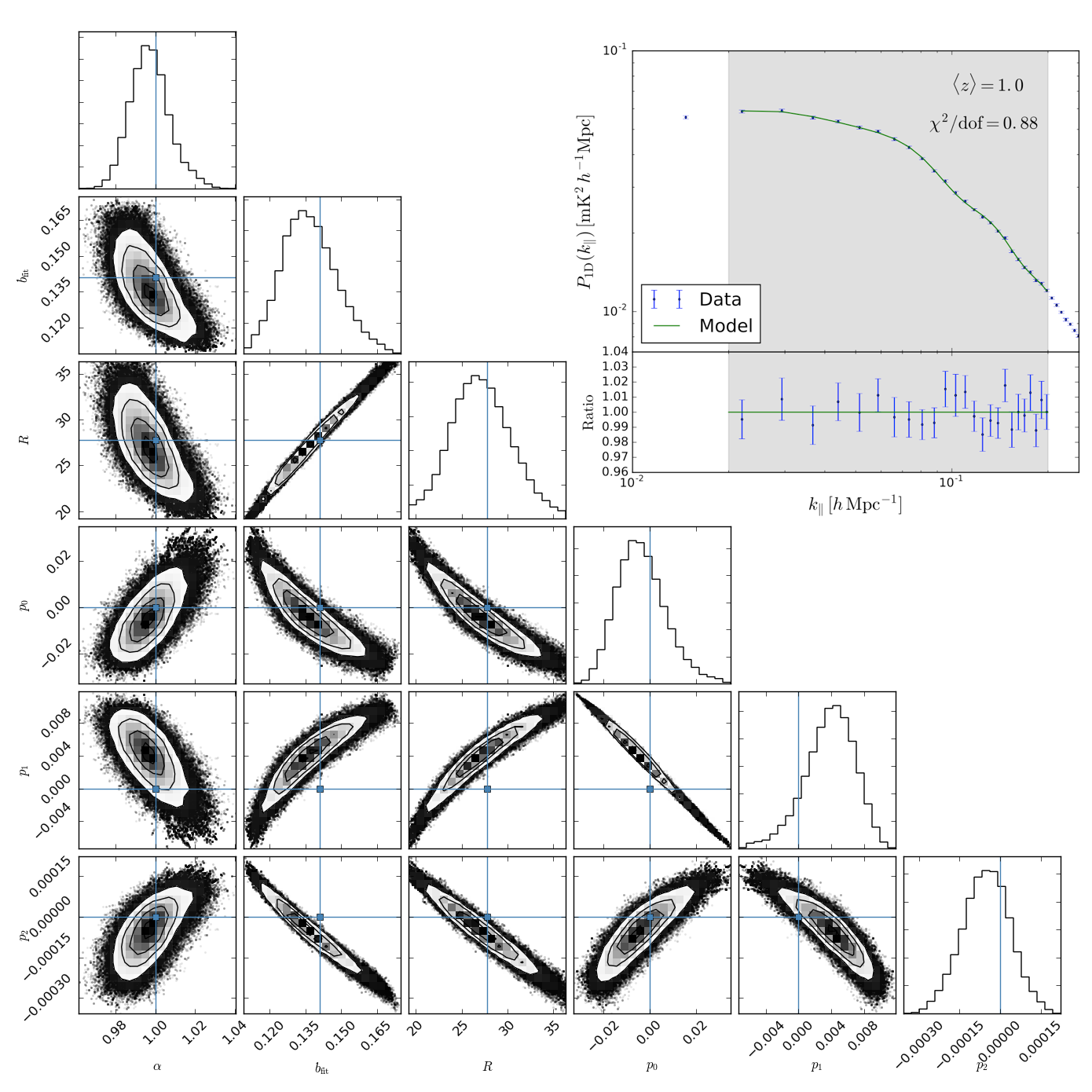}
  \caption{MCMC constraints on the 6 parameters used in the fits to the measured 1D power
           spectrum for one of the simulations. The results of the simulation (at $\langle z \rangle=1.0$)
           together with the best 
           fit is shown in the upper-left corner, where the dashed area represents the $k$-range where
           we perform the fit (residuals shown in the bottom panel). While the BAO parameter
           $\alpha$ shows no significant correlations with the other parameters, a result
           of the robustness of the BAO signal, the remaining five parameters show strong
           degeneracies among themselves. This is due to the fact that all of them affect
           the amplitude and broadband shape of the power spectrum in a similar way.}
  \label{fig:corner}
\end{figure*}

\begin{table*}
\centering
{\renewcommand{\arraystretch}{1.2}
\resizebox{14cm}{!}{
\begin{tabular}{cccccc}
\hline \hline
z range & $\langle z \rangle$ & mask & & $\sigma_\alpha$ &\\ 
 & & & (C) & (C+N) & (C+N+FG)\\
\hline
\multirow{2}{*}{[0.36-0.75]} & \multirow{2}{*}{0.6} & \textcolor{gray}{no} & $\textcolor{gray}{1.008\pm0.016}$ & $\textcolor{gray}{1.008\pm0.016}$ & $\textcolor{gray}{1.007\pm0.016}$ \\
&& \textcolor{black}{yes} & $\textcolor{red}{1.006\pm0.020}$ & $\textcolor{red}{1.006\pm0.021}$ & $\textcolor{blue}{1.006\pm0.024}$ \\
\hline
\multirow{2}{*}{[0.75-1.26]} & \multirow{2}{*}{1.0} & \textcolor{gray}{no} & $\textcolor{gray}{0.996\pm0.010}$ & $\textcolor{gray}{0.997\pm0.011}$ & $\textcolor{gray}{0.996\pm0.011}$ \\
&& \textcolor{black}{yes} & $\textcolor{red}{0.997\pm0.012}$ & $\textcolor{red}{0.997\pm0.013}$ & $\textcolor{blue}{0.998\pm0.015}$ \\
\hline
\multirow{2}{*}{[1.26-1.98]} & \multirow{2}{*}{1.6} & \textcolor{gray}{no} & $\textcolor{gray}{1.001\pm0.011}$ & $\textcolor{gray}{1.004\pm0.014}$ & $\textcolor{gray}{1.003\pm0.014}$ \\
&& \textcolor{black}{yes} & $\textcolor{red}{1.000\pm0.013}$ & $\textcolor{red}{1.003\pm0.016}$ & $\textcolor{blue}{1.004\pm0.019}$ \\
\hline
\multirow{2}{*}{[1.98-3.05]} & \multirow{2}{*}{2.5} & \textcolor{gray}{no} & $\textcolor{gray}{1.004\pm0.013}$ & $\textcolor{gray}{1.003\pm0.021}$ & $\textcolor{gray}{1.000\pm0.021}$ \\
&& \textcolor{black}{yes} & $\textcolor{red}{1.004\pm0.016}$ & $\textcolor{red}{1.002\pm0.026}$ & $\textcolor{blue}{1.002\pm0.031}$ \\
\hline\hline
\end{tabular}}}
\caption{For each simulation we fit the measured radial power spectrum on 4 the redshift bins
         specified in the first column using the theoretical template of Eq. \ref{template}.
         Columns 4-6 show the mean and standard deviation of the BAO parameter $\alpha$.
         Constraints are shown for maps containing only the cosmological signal (column 4),
         the cosmological signal plus instrumental noise (column 5) and the cosmological signal,
         noise and foreground residuals (column 6). The radial power spectra are measured
         using all pixels of the maps (gray numbers) or using a mask: \textcolor{red}{the cosmological mask}
         ($f_{\rm sky}=0.72$) and the \textcolor{blue}{the foregrounds mask} ($f_{\rm sky}=0.58$).}
\label{tbl:best_fit}
\end{table*}

In Fig. \ref{fig:signal-to-noise} we show the signal-to-noise ratio of the radial 21cm power spectrum
in the four redshift bins as a function of wavenumber. The curves shown in this figure were computed
using the theoretical prediction for the 1D power spectrum and its uncertainty assuming Gaussian
statistics, which we have shown to be good approximations to the simulated data (see Appendix
\ref{sec:appendixB}). We considered separately the cases with and without instrumental noise,
and in both cases we estimated the errors assuming the volume available for the cosmological mask.

Focusing on the results involving the cosmological signal alone, it can be seen that the $S/N$
decreases towards higher redshifts. This is a-priori surprising, since the volume of the redshift
bins increases with its mean redshift. We can understand this by writing the $S/N$ ratio at linear
order (see Appendices \ref{sec:noise} and \ref{sec:appendixB} for the derivation) 
\be
{\rm S/N}(k)=\cfrac{\frac{1}{2\pi}\int_0^\infty P_{\rm 21cm,3D,obs}(k_\|,k_\bot)k_\bot dk_\bot}
{\sqrt{(V\Delta k)^{-1}\int_0^\infty
\left[P_{\rm 21cm,3D,obs}(k_\|,k_\bot)+P_N\right]^2k_\bot dk_\bot}}
\ee
where $V$ is the survey volume and $\Delta k$ is the width of the $k$-bin over which the power
spectrum is estimated. $P_N$ is the amplitude of the white-noise arising from the
instrument temperature and is given in Eq. \ref{eq:noise3}.

Should all 21cm maps have the same angular resolution and no system noise, the $S/N$ ratio would simply
scale inversely with the square root of the survey volume. However, the beam size grows at larger
wavelenghts, and the corresponding damping exponential factor $\exp(-k_\bot^2R^2)$ reduces both the
amplitude and the errors of the 1D power spectrum (see Figs. \ref{fig:system_noise} and
\ref{fig:errors_th}) with increasing $R$ (and therefore redshift). The effect on both quantities is,
however, different, and the amplitude of the power spectrum is reduced more efficiently than its
errors. That decrease is not compensated by the increase in the survey volume, and the net effect
is a reduction in $S/N$ at higher redshifts.

Turning now to maps containing instrumental noise, we can see that, while at low redshift
the effects of the noise are almost negligible, in the two highest redshift bins the uncertainties
in the power spectrum become dominated by it. This, for instance, produces a dramatic drop in $S/N$
in the $\langle z\rangle=2.5$ redshift bin. This has a direct impact in the final BAO constraints,
as we shall see in Section \ref{subsec:noise}.

It is worth pointing out that the $S/N$ ratio varies very mildly with wavenumber for the 1D
power spectrum in the noiseless case. This is easy to understand: unlike in the case of the isotropic
3D power spectrum, where each $k$-mode is sampled by all ${\bf k}$s inside a spherical shell of radius
$k$, in the 1D case only a single mode per pixel contributes to the estimate of the power spectrum at
a given $k_\|$, and therefore all modes are sampled roughly equally.

\subsection{Cosmological signal}
\label{subsec:cosmo_signal}
\begin{figure*}
  \centering
  \includegraphics[width=0.95\textwidth]{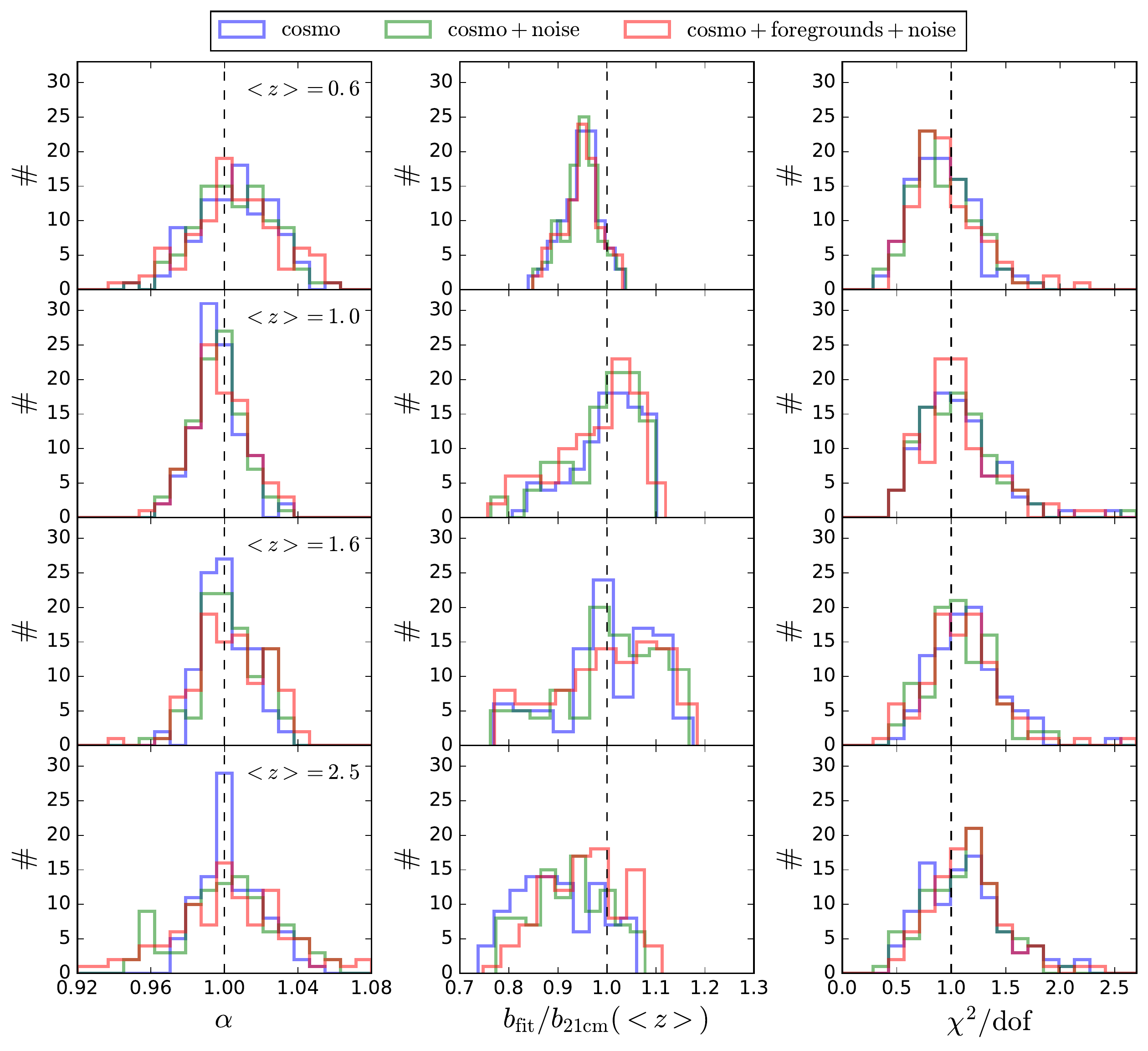}
  \caption{Distribution of the best-fit values of $\alpha$ (left column), $b_{\rm fit}$
           (central column) and the corresponding reduced $\chi^2$ (right column)
           for each redshift bin and simulation. Results are shown for the bins with mean
           redshifts $\langle z\rangle=0.6$ (top), $\langle z\rangle=1.0$ (middle top),
           $\langle z\rangle=1.6$ (middle bottom) and $\langle z\rangle=2.5$ (bottom).
           The different line colors correspond to the results for simulations containing
           only the cosmological signal (blue), cosmological signal plus instrument noise
           (green) and cosmological signal, instrument noise and foreground residuals (red).}
  \label{fig:histo}
\end{figure*}

For simulations containing only the cosmological signal we fit the measured radial 21cm power spectrum
of each redshift bin of each simulation using the template of Eq. \ref{template}. An example of
the measurements together with the best-fit is shown in the upper-right corner of Fig. \ref{fig:corner}. 
From each fit we
obtain the value of the BAO parameter $\alpha$ as well as the nuisance parameters together with
the corresponding $\chi^2$. Fig. \ref{fig:histo} shows, with blue lines, the distribution of
$\alpha$, $b_{\rm fit}$ and reduced $\chi^2$, and the fourth column in Table \ref{tbl:best_fit} shows
the average value and standard deviation of $\alpha$ obtained from the 100 lognormal realizations.
The values obtained for the full sky are displayed in gray, while those corresponding to the
cosmological mask are shown in red. We will focus on the latter from now on.

By inspecting the distribution of the best-fit values of $\alpha$ as well as the corresponding
reduced $\chi^2$ we first verified that our theoretical template is a good model for the data,
and that we obtain an unbiased estimate of the true BAO scale. The relative constraints
on the Hubble rate are given directly by the error on $\alpha$. In the three high-redshift bins we
find $\sigma_\alpha=1.2\%-1.6\%$, while the lowest redshift bin shows a significantly larger error
($\sigma_\alpha=2.0\%$). The constraints on $\alpha$ improve by $\sim20\%$ when using the
full sky, in agreement with the corresponding increase in sky area and therefore survey volume.

The magnitude of the errors on $\alpha$ and its redshift dependence can be understood taking the
two effects discussed in Sections \ref{subsec:radial_BAO} and \ref{subsec:significance}: the $S/N$
ratio of the signal and the significance of the BAO wiggles. While the $S/N$ decreases with redshift
(see Fig. \ref{fig:signal-to-noise}), the significance of the BAO increases with it (see Fig.
\ref{fig:Pk_1D_beam}). The BAO wiggles are less significant on the radial power spectrum at low
redshift, which increases the uncertainties on $\alpha$, and therefore we expect these uncertainties
to increase towards lower redshifts. On the other hand, the significance of the BAO signal
saturates at high redshift, while the $S/N$ ratio continues to decrease with redshift, and thus
we would expect the error on $\alpha$ to grow at higher redshifts as well. This agrees with the
trend observed in the data. The most precise determination of $\alpha$ is obtained in the
two intermediate redshift bins.

Before moving on to more realistic simulations, it is worth noting that, while the parameter
$b_{\rm fit}$ can in principle be interpreted as the bias of the 21cm signal
($b_{\rm fit}\propto b_{\rm HI}\Omega_{\rm HI}$), it is not clear that the method used in
this paper would yield an unbiased estimate of this quantity. The main reason for this is the
strong degeneracies existing between this parameter and the nuisance parameters $R$ and $p_i$
(see Fig. \ref{fig:corner}), all of which affect the overall normalization and broad-band shape
of the power spectrum. Furthermore, unlike the case of the BAO signature, nothing would prevent
us from using the full three-dimensional clustering information to constrain this parameter, and
therefore using the 1D power spectrum to measure $b_{\rm HI}$ would be, in any case, sub-optimal.
Nevertheless, the histograms shown in the central panels of Figure \ref{fig:histo} show that
the recovered values of $b_{\rm 21cm}$ are, on average, in good agreement with the input model.

\subsection{System noise}
\label{subsec:noise}
We now study the impact of the instrument temperature on our results. As shown in Appendix
\ref{sec:noise}, the noise contribution to the total power spectrum increases with redshift,
and in fact, eventually dominates the cosmological signal at low frequencies. We can therefore
expect a non-negligible effect on the detectability and uncertainties of the BAO signal.

\begin{figure}
  \centering
  \includegraphics[width=0.49\textwidth]{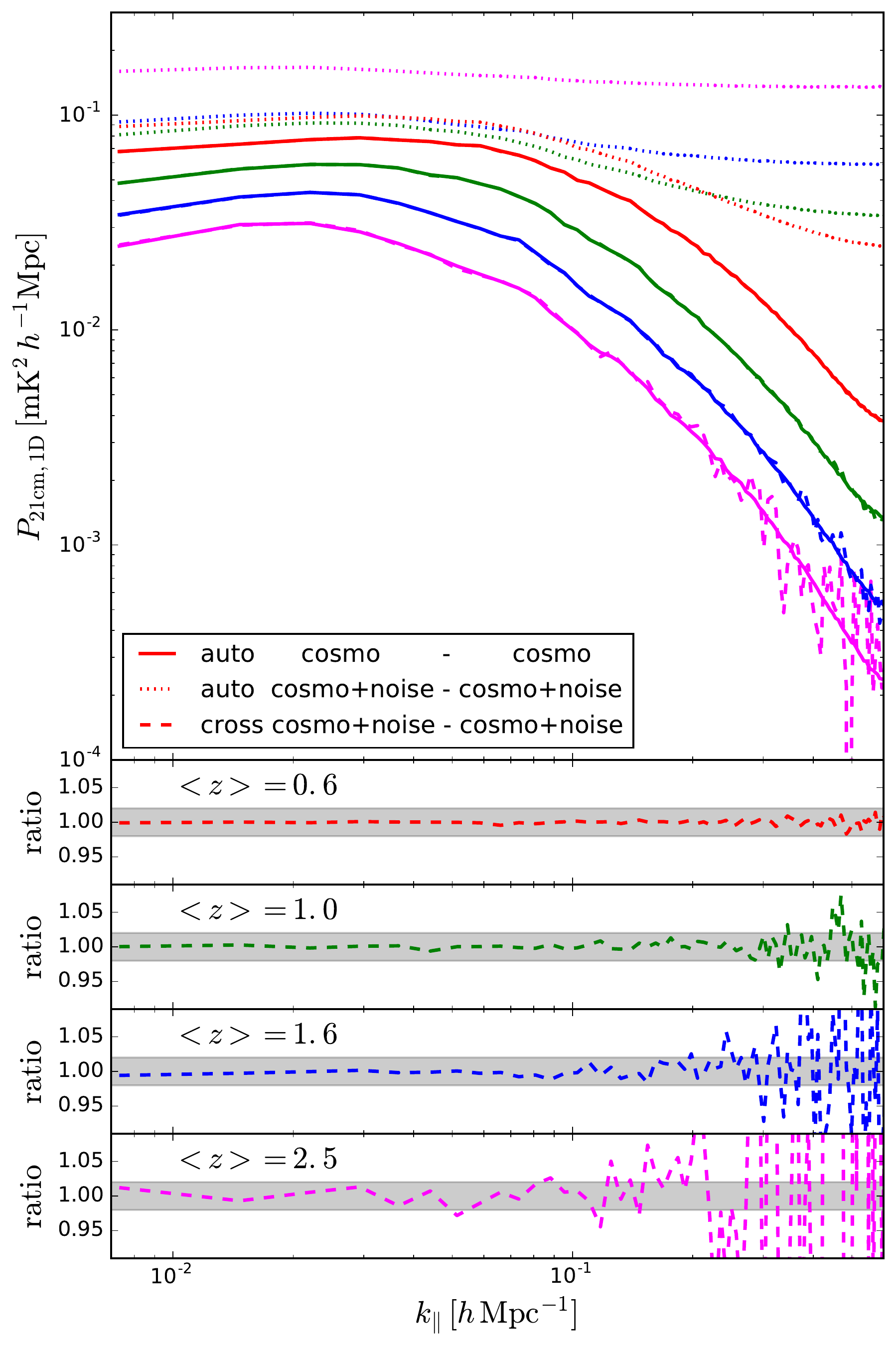}
  \caption{Impact of system noise on the measured radial 21cm power spectrum. The solid
           lines show the power spectrum measured in one simulation containing only
           the cosmological signal. The dotted lines display the results obtained by
           measuring the power spectrum in maps that contain both the cosmological
           signal and the system noise. The dashed lines present the results obtained
           by computing the cross-power spectrum from two different noise realizations
           with the same cosmological signal. The power spectra are measured in four
           different redshift bins: $\langle z\rangle=0.6$ (purple),
           $\langle z\rangle=1.0$ (blue), $\langle z\rangle=1.6$ (green),
           $\langle z\rangle=2.5$ (red). The bottom panels show the ratio between
           dashed and solid lines and shaded regions represent a $2\%$ difference.}
  \label{fig:system_noise}
\end{figure}

In order to avoid a noise bias in the computation of the 1D power spectrum, for each
sky simulation we generate two realizations of the instrumental noise, add them to the
simulation and compute the 1D power spectrum as the cross-correlation of the two 
resulting sets of maps. This simulates the way in which power spectra would be estimated 
in a realistic setting, by taking cross-correlations of different data splits. Fig.
\ref{fig:system_noise} shows the 1D power spectra computed in the four redshift bins of
one of our simulations. The solid lines show the power spectrum computed for a simulation
containing only the cosmological signal, while the dashed lines include the effects of
instrumental noise, which can be observed as a random scatter around the noiseless lines
with a relative variance that grows towards smaller scales. The dotted lines in Figure
\ref{fig:system_noise} illustrate the noise bias induced when auto-correlating maps with
the same noise realization, and explicitly shows the growing contribution of the
instrumental noise towards higher redshifts. This was also illustrated in Figure
\ref{fig:signal-to-noise}.

For each simulation and redshift bin we have fit the radial 21cm power spectrum measured
employing the above procedure to the theoretical template of Eq. \ref{template}. From
each fit we measure the value of $\alpha$ and $\chi^2$ and in Fig. \ref{fig:histo} we
show their distribution in green lines from the 100 realizations for the four different
redshift bins. Table \ref{tbl:best_fit} summarizes, in the fifth column, the mean and
standard deviation of the distribution. We find that the instrumental noise increases the
uncertainty on the BAO parameter by $5\%, 8\%, 23\%\, \text{and}\, 64\%$ in the bins
with mean redshift $0.6, 1.0, 1.6\,\text{and}\,2.5$ respectively. The large degradation
in the BAO signal in the higher redshift bin is, as explained above, a consequence of
the larger instrumental noise present in that bin. It is worth noting that the uncertainties
on the overall scaling of the power spectrum, given by the effective bias $b_{\rm fit}$
experience a much milder variation after introducing instrumental noise. This is due
to the fact that most of the constraining power for this parameter comes from the largest 
scales, where cosmic variance dominates in all bins.

Table \ref{tbl:best_fit} also shows, in gray, the mean values and standard deviations
for full-sky observations. We should clarify that, in this case, we fixed the observing
time per pixel, thus assuming that the total allocated observation time would scale with
$f_{\rm sky}$. The improvement in the constraints is thus solely due to the increase
in surveyed volume.

\subsection{Foregrounds}
\label{subsec:foregrounds}

\begin{figure}
  \centering
  \includegraphics[width=0.49\textwidth]{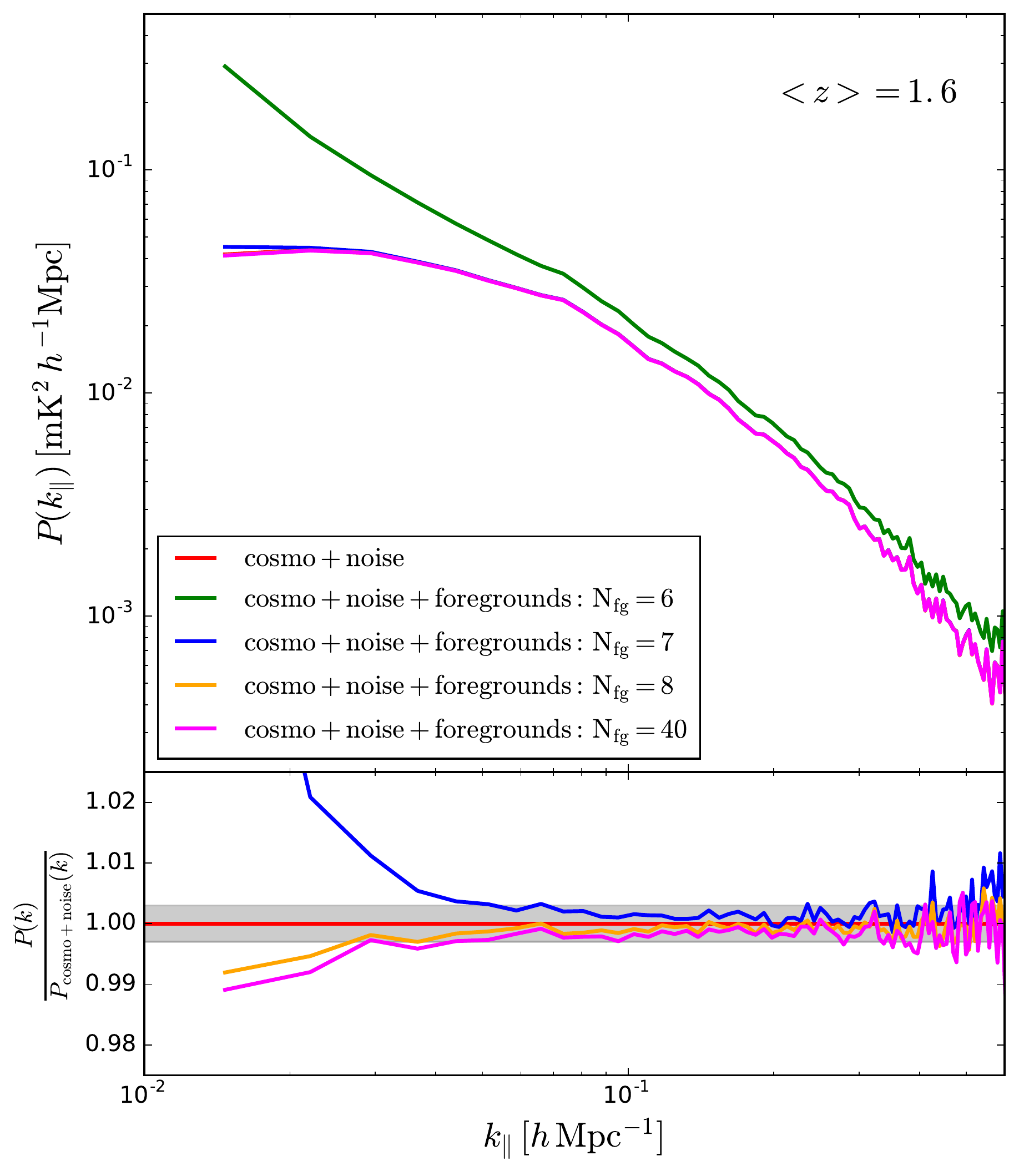}
  \caption{Radial 21cm power spectrum after foreground cleaning for one particular simulation
           in the third redshift bin $\langle z\rangle=1.6$. The green, blue, orange and
           magenta lines correspond to the result of applying a PCA algorithm removing
           $N_{\rm fg}=6$, 7, 8 and 40 principal components. The red solid line shows the
           result for simulations containing only cosmological signal and instrumental noise.
           The bottom panel shows the ratio of the foreground-cleaned power spectra to the result
           of the foreground-free simulation.}
  \label{fig:foregrounds}
\end{figure}

We now study the impact of the presence of Galactic and extra-Galactic foregrounds on our results. 
We added simulations of the most relevant radio foreground sources to each sky realization as
described in Section \ref{subsec:sims} before applying the instrumental beam and noise.
As can be seen in Fig. \ref{fig:mask} the amplitude of the foregrounds is several orders of
magnitude higher than the one from the cosmological signal. We then applied a PCA blind cleaning
algorithm to each simulation as described in Section \ref{subsec:fg} (this was done independently
for the two noise realizations per simulation described in the previous section).

Figure \ref{fig:foregrounds} shows the performance of the foreground cleaning method. The method
is based on subtracting a number $N_{\rm fg}$ of principal components from the maps, with the hope
that the remaining intensity is dominated by the cosmological signal. The figure shows the radial
power spectrum in the third redshift bin ($\langle z\rangle=1.6$) for one particular sky simulation
after having applied a PCA algorithm with different values of $N_{\rm fg}$ ($N_{\rm fg}=6$ in
green, 7 in blue, 8 in orange and 40 in magenta). For comparison the figure also contains, in red,
the power spectrum of the foregrounds-free simulation. It is worth noting that, as described 
in \cite{2015MNRAS.454.3240B}, the effects of correlated instrumental noise, of particular relevance
for single-dish experiments such as SKA1-MID \citep{2016arXiv160609584H}, would also be partially
removed by the foreground-cleaning algorithm. As the figure shows, the procedure works remarkably
well due to the very different spectral characteristics of signal and foregrounds. We find that
the presence of foreground residuals can be minimized after subtracting $N_{\rm fg}=8$ principal
components. Note that, inevitably, this method removes part of the cosmological signal, particularly
on the largest radial scales dominated by foregrounds. This causes a bias in the radial power
spectrum that, however small, could potentially affect the recovery of the BAO scale. This
scale-dependent bias is evident in the lower panel of Figure \ref{fig:foregrounds}, which shows
the ratio of the recovered power spectra with respect to the foreground-less case.

Using the above procedure we calculated the radial 21cm power spectrum in each redshift and simulation
and we fit the results using the template of Eq. \ref{template}. We note that in this case we applied
the optimal Galactic mask described in Section \ref{subsec:sims}, corresponding to sky fraction
$f_{\rm sky}=0.58$. Fig. \ref{fig:histo} shows, in red, the distribution of the fit parameters
$\alpha$ and $b_{\rm fit}$, as well as the values of the reduced $\chi^2$ for our 100 simulations.
The corresponding mean value and standard deviation of $\alpha$ for simulations containing foregrounds
are reported in the sixth column of Table \ref{tbl:best_fit}. As expected, we find that the errors on
$\alpha$ increase with respect to the results found in the foreground-free simulations. The error
enhancement is roughly $\sim20\%$, and is mainly caused by the smaller sky area allowed by the Galactic
mask. More importantly, we find that as expected, the small foreground bias on large scales mentioned
above does not bias the recovered values of the BAO scaling parameter. Note that, as pointed out in
\cite{2015MNRAS.447..400A}, the power spectrum uncertainties are also expected to receive a contribution
from foreground residuals, however, this effect is subdominant. We thus conclude that, for well-behaved
foregrounds, SKA1-MID should be able to measure the radial BAO scale with good accuracy up to redshifts
$z\sim3$.

\subsection{BAO significance}
\label{subsec:significance}
Besides placing constraints on the scale of the BAO, it is also important to determine the significance
with which the signature has been detected in the power spectrum.
\begin{figure*}
  \centering
  \includegraphics[width=0.99\textwidth]{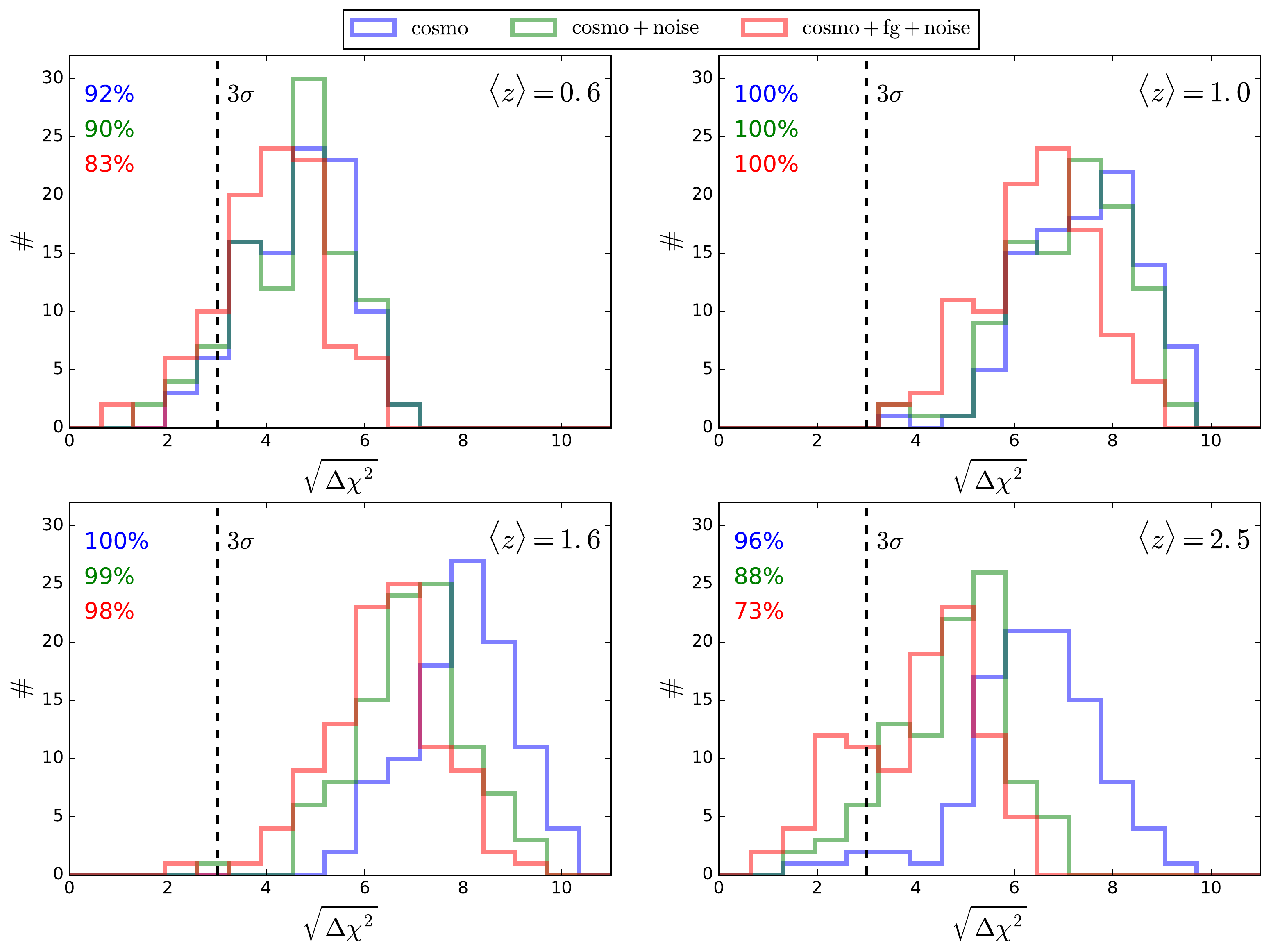}
  \caption{Significance of the BAO wiggles in each redshift bin for maps containing only the
           cosmological signal (blue), the cosmological signal plus system noise (green) and the
           cosmological signal, noise and foregrounds (red). The histograms show the distribution
           of the square root of the increment in $\chi^2$ between our fiducial model for the
           radial power spectrum (Eq. \ref{template}) and a no-BAO template. Thus the $x$-axis
           can be read as the ``number of sigmas'' with which the BAO have been detected. The
           vertical dashed lines show the $3\sigma$ threshold, and we provide, in each case, the
           number of simulations with a significance above this threshold.}
  \label{fig:BAO_significance}
\end{figure*}
We have done so through the following procedure. For each simulation and redshift bin we measure the
radial 21cm power spectrum and then fit the results using a no-BAO template for the power spectrum:
\be
P_{\rm model}(k_\|,z|\vec{\Theta})=P_{\rm nw,1D}(k_\|,z)+p_0k_\|+p_1+p_2/k_\|~.
\ee
We then compared the $\chi^2$ value of the fits with and without the BAO contribution, and quantified
the significance of the BAO detection in terms of the increment between both cases,
$\Delta \chi^2$.  Fig. \ref{fig:BAO_significance} shows the histograms for the values of
$\sqrt{\Delta\chi^2}$ from each redshift bin and simulation, for maps containing just the
cosmological signal (blue lines), the cosmological signal plus system noise (green lines) and the
cosmological signal, noise and foreground residuals (red lines). 

\begin{figure}
  \centering
  \includegraphics[width=0.49\textwidth]{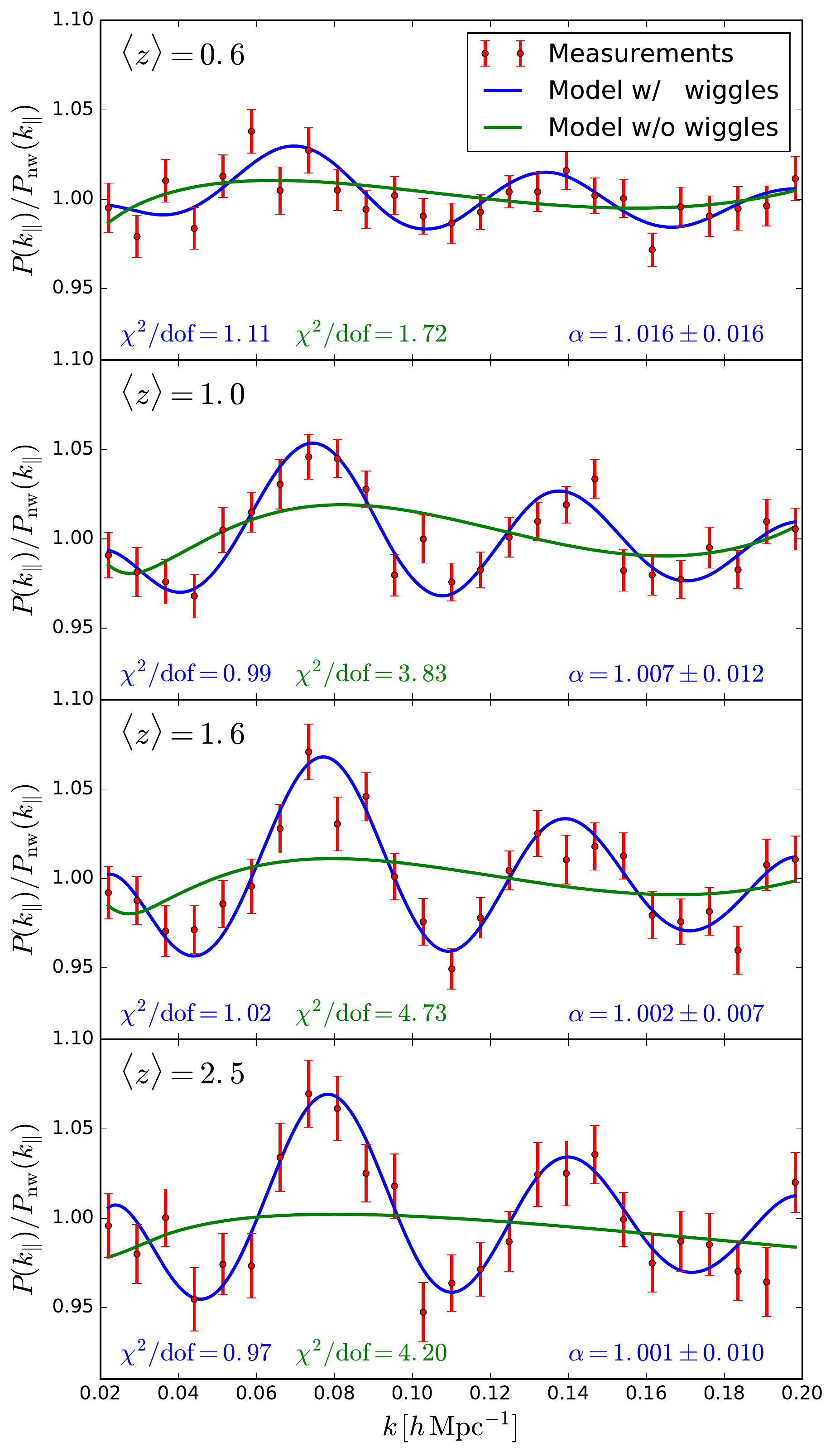}
  \caption{The red points represent measurements of the radial power spectrum in different
           redshift bins from one particular simulation. The error bars of the red points
           represent the error on the measurements. The blue/green lines display the best-fit
           model from a theoretical template with/without BAO wiggles. Both the measurements
           and the best-fit models are normalized by a model without BAO wiggles with fixed
           parameters.}
  \label{fig:fit}
\end{figure}

We find that, although in the higher-redshift bin a relatively large fraction ($\sim27\%$) of
the simulations yield BAO detections below the $3\sigma$ threshold, it is generally likely, in
all cases, to make a reliable measurement ($>3\sigma$) of the radial BAO signature. As explained
above, the main reason for the lower significance of the BAO measurements in the highest redshift
bin is the larger system noise at low frequencies. It is also worth noting that the first redshift
bin generally yields lower-significance detections than the next two. As discussed in Section
\ref{subsec:radial_BAO}, this is caused by the larger contribution from small angular scales in
the lower redshift bin, which reduce the relative contribution from the BAO wiggles to the radial
power spectrum. This can be visualized explicitly in Fig. \ref{fig:fit}, which shows the best-fit
power spectra with and without BAO in the 4 different redshift bins for one particular simulation
in the absence of noise or foreground contamination.

\section{Conclusions}
\label{sec:conclusions}
The BAO scale is one of the most robust cosmological observables, due to the distinctive
nature of their signature, a single peak on the 2pt correlation function or a set of wiggles
on the power spectrum. This observable can be used to measure the Hubble function and the
angular diameter distance as a function of redshift, and therefore represents a unique and
robust probe to study  the nature of dark energy.  

The purpose of this paper has been to investigate the accuracy with which the BAO scale can
be determined through single-dish 21cm intensity mapping observations in the post-reionization
epoch. We quote results for a possible intensity mapping experiment carried out with the
SKA1-MID array covering more than half of the sky in the redshift range $z\sim[0.3-3]$.
We however emphasize that our methodology is fully general and can be easily applied to
other instruments. 

We have shown that the smearing caused by the beam size of the radio-telescopes will prevent
a competitive measurement of the isotropic BAO scale in both the 21cm correlation function
or power spectrum (see Fig. \ref{fig:beam_BAO}). However, we have shown that, given the good
frequency resolution of radio telescopes, it should still be possible to measure the radial
BAO signal down to high redshifts, thus placing competitive constraints on the expansion
rate $H(z)$. 

In this paper we have proposed a method to recover the radial BAO scale in intensity mapping
observations and implemented it in practice making use of a suite of 100 full-sky lightcone
simulations in order to systematically study the effects of the instrumental noise and the
robustness of the signal to foreground-related systematic effects. Our procedure follows
three steps:
\begin{itemize}
  \item A simulated sky is generated containing a realization of the full-sky HI cosmological
        signal as well as the most relevant Galactic and extra-Galactic foregrounds in the
        frequency range $\nu\in[350,1050]\,{\rm MHz}$. The simulated maps are smoothed to the
        angular resolution corresponding to the specifications of SKA1-MID, and white
        instrumental noise is added accordingly.
  \item We remove the foregrounds using a PCA algorithm, subtracting the first 8 principal
        components, which we have shown are dominated by foregrounds.
  \item We compute the radial power spectrum of the resulting maps by stacking the 1-dimensional
        Fourier transform of every pixel in the field of view along the frequency direction
        (further details about the method are given in Section \ref{subsec:pk}).
  \item For each estimated power spectrum, we determine the radial BAO scaling parameter
        $\alpha$ by fitting the template given in Eq. \ref{template} to the data. The mean value
        and uncertainty on $\alpha$ is then estimated by averaging over 100 simulations.
\end{itemize}
All our simulations take into account the sky area limitations of the SKA both in terms
of accessible sky and Galactic foregrounds.

Our results concerning the measurement of the BAO scale are summarized in Table
\ref{tbl:best_fit}. We find that the BAO uncertainties become larger at both high
and low redshifts, even in the absence of instrumental noise. We have shown that
this is due to the low relative amplitude of the BAO signature in the radial 
power spectrum at low redshifts and to the lower signal-to-noise ratio of the
total HI power spectrum at high redshifts caused by the larger size of the telescope
beam, which overcomes the $\propto1/\sqrt{V}$ improvement factor due to the larger
volume coverage. More importantly, we find that, while the effects of instrumental
noise are irrelevant at low redshift, they come to dominate the error budget at
redshifts $z\gtrsim2$, increasing the final BAO uncertainties by a factor of $\sim2$
with respect to the sample-variance limited result.

Concerning the effect of radio foregrounds, we have shown that the large-scale
bias induced by foreground removal on the radial power spectrum (as reported by
e.g. \cite{2015MNRAS.447..400A}) does not cause a bias in the recovered BAO scale.
This is thanks to the robustness of the BAO signal against broad-band variations
in the shape of the power spectrum, as well as to the spectral separation between
foregrounds and cosmological signal in the frequency direction. We have also shown
that the contribution of foreground residuals to the final uncertainties is
negligible, and that, therefore, the main effect of foregrounds is a reduction in
the available sky area needed in order to avoid the regions of higher Galactic
emission. Although we have not explicitly introduced this effect, it should be
possible to mitigate the impact of correlated instrumental noise, of particular
relevance to single-dish observations, using similar methods
\citep{2015MNRAS.454.3240B}.

Finally, we have studied the significance of the BAO signal. We have shown that,
although the large variance of the instrumental noise at low frequencies reduces 
the signal-to-noise ratio of the BAO signature at high redshifts, we obtain 
significant detections ($>3\sigma$) of it in a large majority of our simulations
in all redshift bins.

Overall, we conclude that by a single-dish 21cm intensity mapping experiment carried
out by SKA1-MID over $\sim50\%$ of the sky with an allocated observing time of 10000
hours should be able to place direct constrains the value of the Hubble function $H(z)$
with a relative uncertainty of $(2.4\%, 1.5 \%, 1.9\%, 3.1\%)$ at redshifts
$\langle z\rangle=(0.6,1.0,1.6,2.5)$ by measuring the BAO scale in the radial 21cm
power spectrum. This would correspond to a precision comparable with next-generation
spectroscopic surveys (e.g. \citet{2014JCAP...05..023F}).

\section*{Acknowledgements} 
We thank Andrej Obuljen and Mario Santos for useful comments and conversations.
FVN and MV are supported by the ERC Starting Grant ``cosmoIGM'' and partially
supported by INFN IS PD51 ``INDARK''. DA is supported by the Beecroft Trust
and ERC grant 259505. We acknowledge partial support from ``Consorzio per la
Fisica - Trieste''.

\begin{appendix}

\section{HI model}
\label{sec:appendix_a}

It has been shown in \cite{Villaescusa-Navarro_2014a} that the abundance of HI
outside dark matter halos is negligible and that its contribution to the
amplitude of the 21cm power spectrum can be safely neglected. Under those
conditions, it is possible to model the clustering properties of HI using the
halo model framework. A key element in that formalism is the function
$M_{\rm HI}(M,z)$, that outputs the average HI mass that a dark matter halo
of mass $M$ has at redshift $z$. If this  function is known, it is possible
to compute the two basic elements needed to estimate the shape and amplitude
of the 21cm power spectrum at linear order, $\Omega_{\rm HI}(z)$ and $b_{\rm HI}$:
\begin{eqnarray}
\Omega_{\rm HI}(z)&=&\frac{1}{\rho_c^0}\int_0^\infty n(M,z)M_{\rm HI}(M,z)dM\\
\nonumber\\
b_{\rm HI}(z)&=&\frac{1}{\rho_c^0\Omega_{\rm HI}(z)}\int_0^\infty b(M,z)n(M,z)M_{\rm HI}(M,z)dM\nonumber
\end{eqnarray}
where $\rho_c^0$ is the critical density of the Universe today and $n(M,z)$ and
$b(M,z)$ are the halo mass function and halo bias at redshift $z$. 

\cite{Villaescusa-Navarro_2015b} used zoom-in hydrodynamical simulations to show
that the high mass end of the $M_{\rm HI}(M,z)$ can be modeled by a simple power
law: $M_{\rm HI}(M,z)\propto M^{3/4}$. In the low-mass end \cite{Villaescusa-Navarro_2015a}
found a similar behavior for many different cosmological models. We therefore model
the $M_{\rm HI}(M,z)$ function as 
\begin{equation}  
M_{\rm HI}(M,z) = e^{\gamma(z)} M^{3/4}\exp\left(-(M_{\rm min}(z)/M)^2\right)
\label{M_HI_fit}
\end{equation} 
where $e^{\gamma(z)}$ represents an overall normalization and the exponential cut-off
at the characteristic halo mass $M_{\rm min}(z)$ is introduced to model the fact that
it is expected that low-mass halos should not host a significant amount of HI
\citep{Pontzen_2008, Marin_2010, Bagla_2010, Villaescusa-Navarro_2014a, Hamsa_2015b}.
For simplicity we consider that the characteristic cut-off scale $M_{\rm min}(z)$
does not depend on redshift.

Our model has two free parameters: $M_{\rm min}$ and $\gamma(z)$. The value value
of $\gamma(z)$ is fixed by requiring that our model reproduce the relation
$\Omega_{\rm HI}(z)=4\times10^{-4}(1+z)^{0.6}$ inferred from observations in
\cite{Crighton_2015}. The value of $M_{\rm min}=1.7\times10^{10}~h^{-1}M_\odot$
is chosen such as our model reproduces the value of $\Omega_{\rm HI}b_{\rm HI}=
(0.62^{+0.23}_{-0.15})\times10^{-3}$ at $z\simeq0.8$ derived from 21cm intensity
mapping observations in \cite{Switzer_2013}. 

The HI bias that we obtain with the above model can be well described in the redshift range $z\in[0,3]$ by the following relation
\be
b_{\rm HI}(z)=0.904+0.135(1+z)^{1.696}~.
\ee
We notice that our model reproduces, by construction, the $\Omega_{\rm HI}(z)$
relation, while at $z=0.8$ it predicts a value of $\Omega_{\rm HI}b_{\rm HI}=7.2\times10^{-4}$,
in perfect agreement with the constraints from \cite{Switzer_2013}. Besides, at $z=2.3$ 
it predicts a HI bias equal to $b_{\rm HI}=1.93$, compatible with the measured DLAs bias
by \cite{Font_2012}, $b_{\rm DLAs}(z=2.3)=2.17\pm0.2$, assuming that bias of the DLAs is
a good proxy for the HI bias\footnote{This may not be true in many situations, e.g.
Castorina et al. in prep.}.

\section{Noise power spectrum}
\label{sec:noise}
The aim of this appendix is to compute the theoretical noise contribution to the total
power spectrum. We start by noting that an uncorrelated Gaussian random field $n$ has
a white power spectrum given by
\begin{equation}
 \langle\delta_{\bf k}\delta^*_{\bf q}\rangle=\delta^{\cal D}({\bf k}-{\bf q})\,
 v_{\rm cell}\sigma_{\rm cell}^2,
\end{equation}
where $\sigma_{\rm cell}^2$ is the variance of the field in cells of volume $v_{\rm cell}$.

On the other hand, the instrumental noise variance per pixel is given by
\begin{equation}\label{eq:noise1}
 \sigma^2_{\rm pix}=
 \frac{T_{\rm sys}^2}{2\Delta\nu\,t_{\rm pix}\,N_{\rm dish}},
\end{equation}
where $T_{\rm sys}$ is the system temperature, $\Delta\nu$ is the frequency interval, $t_{\rm pix}$ is
the observation time per pixel and $N_{\rm dish}$ is the number of antennas.
$t_{\rm pix}$ can be related to the pixel solid angle as
$t_{\rm pix}=\Omega_{\rm pix}\,t_{\rm tot}/(4\pi\,f_{\rm sky})$, where 
$t_{\rm tot}$ is the total observation time. Moreover, the frequency interval
$\Delta\nu$ and the pixel solid angle $\Omega_{\rm pix}$ can be related to the
comoving volume covered by each pixel as
\begin{equation}\label{eq:noise2}
  \Delta\nu\,\Omega_{\rm pix}=\frac{H\,\nu_{\rm 21}}{r^2\,(1+z)^2}\,v_{\rm pix},
\end{equation}
where $r$ and $H$ are the comoving distance and the expansion rate.

Thus, combining Equations \ref{eq:noise1} and \ref{eq:noise2} we obtain the noise power
spectrum
\begin{equation}\label{eq:noise3}
 P_N=T_{\rm sys}^2
 \frac{4\pi\,f_{\rm sky}[(1+z)\,r]^2}{2\,t_{\rm tot}\nu_{21}N_{\rm dish}H}.
\end{equation}

\section{Covariance matrix}
\label{sec:appendixB}
\begin{figure}
  \centering
  \includegraphics[width=0.49\textwidth]{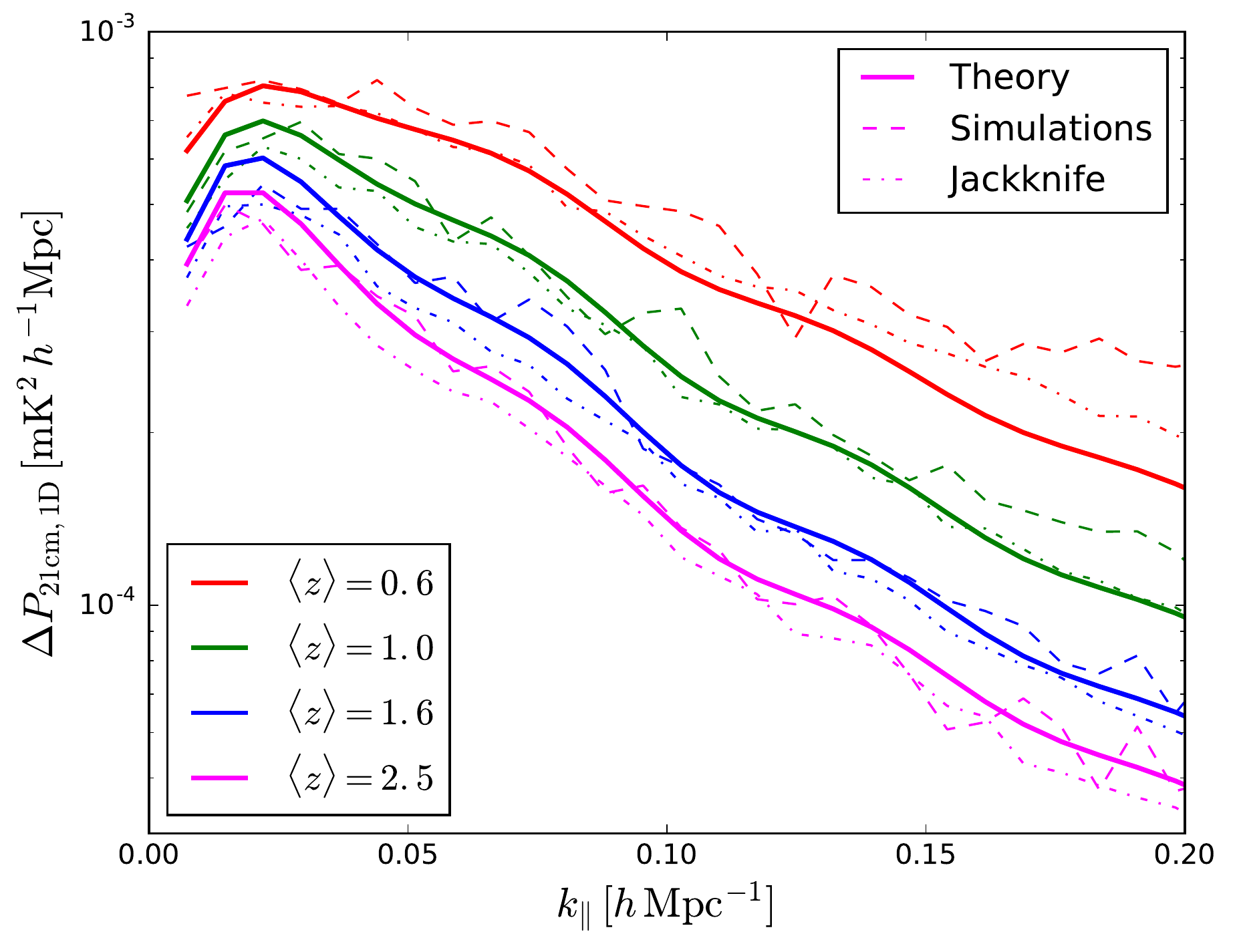}
  \caption{1$\sigma$ uncertainties on the radial 21cm power spectrum estimated from
           the theoretical Gaussian prediction (solid lines), as the standard
           deviation of the 100 lognormal realizations (dashed lines) and using
           the Jackknife technique over one single realization (dot-dashed lines).
           Results are shown for the four fiducial redshift bins we use in our analysis.}
  \label{fig:errors_th}
\end{figure}
\begin{figure}
  \centering
  \includegraphics[width=0.49\textwidth]{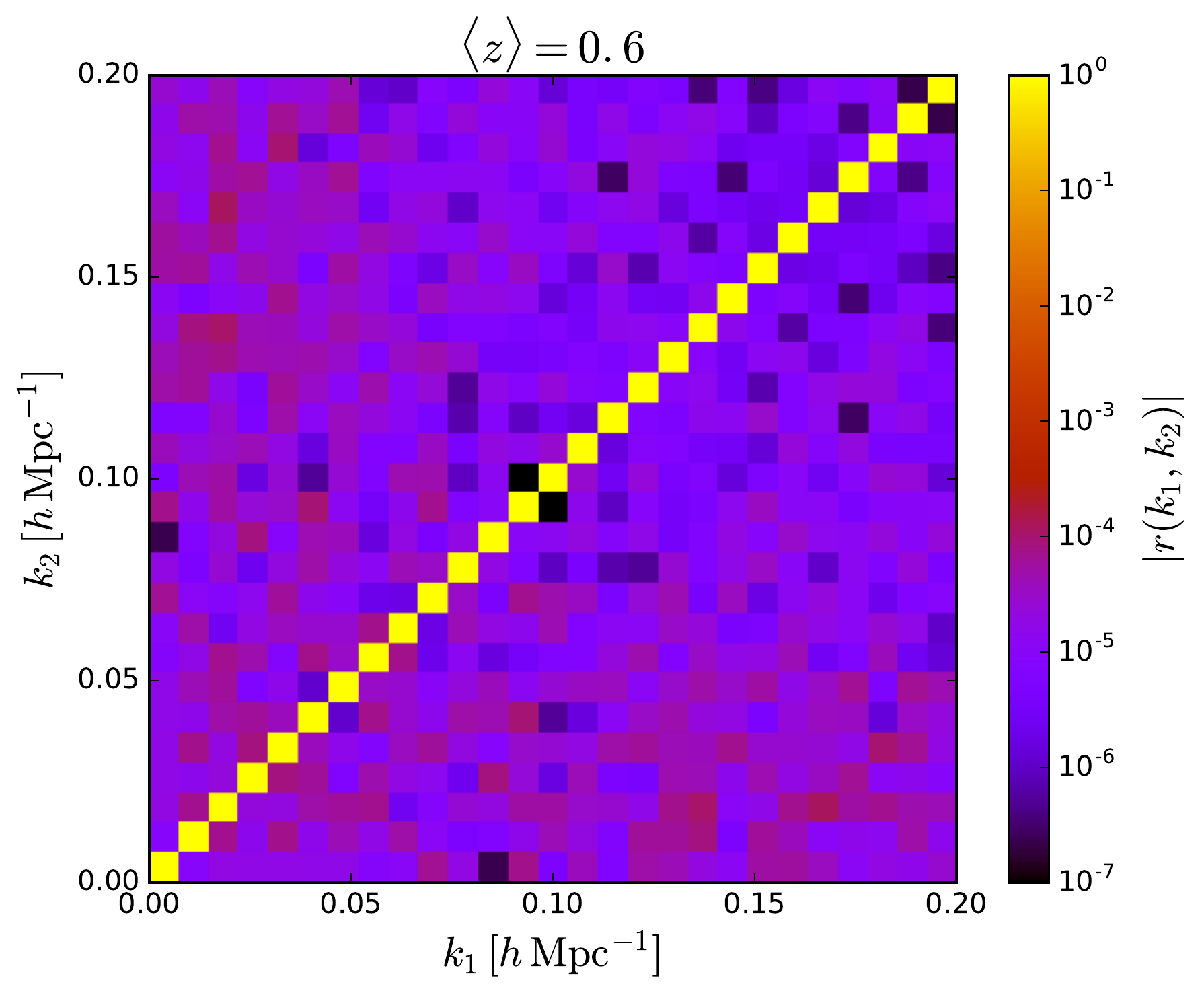}
  \caption{Absolute value of the correlation coefficient matrix of the radial power spectrum
           in the $\langle z\rangle=0.6$ redshift bin estimated from 100 lognormal realizations.}
  \label{fig:covariance}
\end{figure}
In order to validate the uncertainties on the radial power spectrum used in the analysis of this
work we have compared the results from three different methods: 1) the expected Gaussian errors, 2) the r.m.s. from our 100 simulations and 3) the errors estimated using the Jackknife method.

We begin by describing the computation of the expected Gaussian errors. Throughout this Section we will make use of the flat-sky approximation. The basic observable in our analysis is the HI overdensity field Fourier-transformed along the
line of sight (labelled by $\parallel$ here):
\begin{equation}
  \delta(\kpar,{\bf r}_\perp)\equiv\int\frac{dr_\parallel}{\sqrt{2\pi}}\,
  \delta(r_\parallel,{\bf r}_\perp)\,e^{i\kpar r_\parallel}.
\end{equation}
As we described above, the 1D power spectrum is estimated by averaging the radial power
spectrum across all lines of sight independently. Thus, our estimator is:
\begin{equation}
  \hat{P}_{\rm 1D}(\kpar)=\frac{2\pi}{L}\int\frac{d^2r_\perp}{A}
  \left|\delta(\kpar,{\bf r}_\perp)\right|^2,
\end{equation}
where $L$ and $A$ are the comoving depth and area of the region covered by the intensity
mapping experiment, and the factor $2\pi/L$ accounts for the Dirac's delta normalization
of the power spectrum in a region of finite size.

Now, assuming that the HI overdensity is Gaussianly distributed, we can use Wick's
theorem to show that the covariance of this estimator is given by
\begin{align}\nonumber
  C(\kpar,\kpar')&\equiv\left\langle\left(\hat{P}_{\rm 1D}(\kpar)-P_{\rm 1D}(\kpar)\right)
                 \left(\hat{P}_{\rm 1D}(\kpar')-P_{\rm 1D}(\kpar')\right)\right\rangle\\
                 &=\frac{\delta^{\cal D}(\kpar-\kpar')}{V}
                 \int_0^\infty dk_\perp\,k_\perp\,P_{\rm 3D}^2(\kpar,k_\perp),
\end{align}
where $V\equiv L\,A$, and $\delta^{\cal D}$ is the Dirac delta function. Since we
measure the 1D power spectrum in finite intervals of $\kpar$ of width $\Delta\kpar$,
we can substitute $\delta^{\cal D}(\kpar-\kpar')\rightarrow
\delta^{\cal K}(\kpar,\kpar')/\Delta\kpar$, where $\delta^{\cal K}$ is the
Kronecker delta. Thus we find that, in the Gaussian approximation, the uncertainties
on the 1D power spectrum are purely diagonal and given by
\begin{equation}
 {\rm Var}\left[P_{\rm 1D}(\kpar)\right]=\frac{1}{V\Delta\kpar}
          \int_0^\infty dk_\perp\,k_\perp\,P_{\rm 3D}^2(\kpar,k_\perp)~.
\end{equation}

We conclude by noting that, in the presence of noise, $P_{\rm 3D}$ above must be understood
to contain contributions from both the cosmological signal and the instrumental noise.
Taking into account the pixel window function this then reads:
\begin{equation}
  P_{\rm 3D}^2(\kpar,k_\perp)=W^2_p(k_\perp)\left[P_{\rm 21cm}(\kpar,k_\perp)+P_N\right],
\end{equation}
where the noise power spectrum $P_N$ is given in Eq. \ref{eq:noise3}.

Figure \ref{fig:errors_th} shows this theoretical prediction compared with the standard
deviation of the 100 lognormal realizations and the error estimation from one single
realization using the Jackknife method. We show the results for the 4 fiducial frequency bins
used in the BAO analysis. We conclude that the error estimations
from the three different methods are in very good agreement among themselves. We notice
however that there are non-negligible deviations in the highest frequency bin (lowest
redshifts) on small scales (large $\kpar$), which can be ascribed to the effect of the
non-linearities induced by the lognormal transformation. We also find that errors computed
using the Jackknife method tend to be systematically lower than those obtained from the
other two methods. Figure \ref{fig:covariance} shows the absolute value of the
correlation matrix,
\be
r(k_1,k_2)=\frac{C(k_1,k_2)}{\sqrt{C(k_1,k_1)C(k_2,k_2)}}~,
\ee
measured on the most non-linear redshift bin: $z\in[0.36-0.75]$ from the 100 lognormal
realizations. As can be seen, the non-diagonal elements of the covariance matrix are 
negligible, which justifies our use of purely diagonal errors.

We therefore conclude that the covariance matrix of the radial 21cm power spectrum, in the
redshifts and $k-$range relevant for this paper, can be accurately approximated by a diagonal
matrix,  whose elements can be found by employing the theoretical Gaussian prediction, the
standard deviation of different realizations or internal methods such as Jackknife. The results
presented in this papers were obtained using errors computed from the 100 lognormal realizations.

\end{appendix}

\bibliographystyle{mn2e}
\setlength{\bibhang}{2.0em}
\setlength\labelwidth{0.0em}
\bibliography{references}

\end{document}